\documentclass[11pt, notitlepage]{article}

\usepackage{makeidx}
\usepackage{amsmath}
\usepackage{amsfonts}
\usepackage{eucal}
\usepackage[onehalfspacing]{setspace}
\usepackage{geometry}
\usepackage{amssymb}
\usepackage{graphicx}
\usepackage[dvips]{epsfig}
\usepackage{tikz}%
\usepackage{setspace}
\usepackage{natbib}
\usepackage{fancybox}
\usepackage{pstricks}
\usepackage{dsfont}

\tikzstyle{trans} = [line width=2pt,->]
\tikzstyle{transS} = [line width=2pt]

\newcommand{\titledframe}[2]{%
\boxput*(0,1){\psframebox*{#1}}%
{\psframebox[framesep=12pt]{#2}}}

\def\tv{\vrule width 0pt height 20pt depth 5pt}
\def\tvi{\vrule width 0pt height 15pt depth 5pt}

\def\arg{{\rm arg}}

\begin{document}
\newtheorem{thm}{Theorem}
\newtheorem{prop}[thm]{Proposition}
\newtheorem{remark}[thm]{Remark}
\newtheorem{assumption}[thm]{Assumption}

\numberwithin{equation}{section}

\title{\textsc{State-Observation Sampling and the Econometrics of Learning Models}}
\author{\textbf{Laurent E. Calvet and Veronika Czellar}\thanks{Calvet: Department of Finance, HEC Paris, 1 rue de la Lib\'{e}ration, 78351 Jouy en Josas, France; calvet@hec.fr. Czellar: Department of Economics and Decision Sciences, HEC Paris, 1 rue de la Lib\'{e}ration, 78351 Jouy en Josas, France, czellarv@hec.fr.
An earlier version of this paper was circulated under the title ``Efficient Estimation of Learning Models.'' We received helpful comments from Pavel Chigansky, Nicolas Chopin, Adlai Fisher, Thierry Foucault, Itzhak Gilboa,  Christian Gouri\'eroux, Lars Hansen, Per Mykland, Hashem Pesaran, Nicholas Polson, Elvezio Ronchetti, Andrew Siegel, Ramon van Handel, Paolo Zaffaroni, and seminar participants at CORE, the University of Chicago, the University of Evry, the Second HEC Finance and Statistics Conference, CEF 2010 in London, the 2010 Toulouse School of Economics Financial Econometrics Conference, ISBIS 2010 in Portoroz, and the 2010 Econometric Society World Congress in Shanghai. We gratefully acknowledge the computer support of EUROFIDAI and the financial support of the American Statistical Association, the Europlace Institute of Finance, and the National Institute of Statistical Sciences.}}

\date{First version: February 2010\\
This version: May 2011}
\maketitle

\begin{abstract}
\noindent In nonlinear state-space models, sequential learning about the hidden state can proceed by particle filtering when the density of the observation conditional on the state is available analytically (e.g. Gordon {\it et al.} 1993). This condition need not hold in complex environments, such as the incomplete-information equilibrium models considered in financial economics. In this paper, we make two contributions to the learning literature. First, we introduce a new filtering method, the state-observation sampling (SOS) filter, for general state-space models with intractable observation densities. Second, we develop an indirect inference-based estimator for a large class of incomplete-information economies. We demonstrate the good performance of these techniques on an asset pricing model with investor learning applied to over 80 years of daily equity returns.
\end{abstract}
\vskip30pt
\thispagestyle{empty}
 
\newpage
\setcounter{page}{1}
\section{Introduction}
Sequential learning by economic agents is a powerful mechanism that theoretically explains key properties of asset returns, aggregate performance and other equilibrium outcomes (e.g., P\'astor and Veronesi, $2009a$).\footnote{In financial economics, investor learning has been used to explain phenomena as diverse as the level and volatility of equity prices, return predictability, portfolio choice, mutual fund flows, firm profitability following initial public offerings, and the performance of venture capital investments. In particular, the portfolio and pricing implications of learning are investigated in Brennan (1998), Brennan and Xia (2001), Calvet and Fisher (2007), David (1997), Guidolin and Timmermann (2003), Hansen (2007), P\'astor and Veronesi (2009$b$), Timmermann (1993, 1996), and Veronesi (1999, 2000). We refer the reader to P\'astor and Veronesi (2009$a$) for a recent survey of learning in finance.} In order to use these models in practice, for instance to forecast and price assets, a crucial question arises: How can we track agent beliefs? A natural possibility is to consider particle filters, a large class of sequential Monte Carlo methods designed to track a hidden Markov state from a stream of partially revealing observations (e.g. Gordon, Salmond, and Smith, 1993; Johannes and Polson, 2009; Pitt and Shephard, 1999). Existing filtering methods, however, are based on the assumption that the density of the observation conditional on the hidden state (called {\it observation density}) is available in closed form up to a normalizing constant. This assumption is unfortunately not satisfied in incomplete-information economies. In this paper, we introduce the state-observation sampling (SOS) filter, a novel sequential Monte Carlo method for general state space models with intractable observation densities. In addition, we develop an indirect inference-based estimator (Gouri\'eroux, Monfort and Renault 1993; Smith, 1993) for the structural parameters of an incomplete-information economy. 

Since their introduction by Gordon, Salmond, and Smith (1993), particle filters have considerably expanded the range of applications of hidden Markov models and now pervade fields as diverse as engineering, genetics, statistics (Andrieu and Doucet, 2002; Chopin, 2004; Kuensch, 2005), finance (e.g. Kim, Shephard and Chib, 1998; Johannes, Polson, and Stroud, 2009), and macroeconomics (Fern\'andez-Villaverde and Rubio-Ramirez, 2007; Fernandez-Villaverde et al., 2009; Hansen, Polson and Sargent, 2011).\footnote{Advances in particle filtering methodology include Andrieu, Doucet, and Holenstein (2010), Del Moral (2004), Fearnhead and Clifford (2003), Gilks and Berzuini (2001), Godsill, Doucet, and West (2004), and Storvik (2002). Particle filters have received numerous applications in finance, such as model diagnostics (Chib, Nardari, and Shephard, 2002), simulated likelihood estimation (Pitt, 2005), volatility forecasting (Calvet, Fisher, and Thompson, 2006), and derivatives pricing (Christoffersen, Jacobs, and Mimouni 2007). See Capp\'e, Moulines and Ryd\'en (2005), Doucet and Johansen (2008), and Johannes and Polson (2009) for recent reviews.} 
These methods provide estimates of the distribution of a hidden Markov state $s_t$ conditional on a time series of observations $R_t=(r_1,...,r_t)$, $r_t\in\mathbb{R}^{n_R}$, by way of a set of ``particles'' $(s_t^{(1)},...,s_{t}^{(N)}).$ In the original sampling and importance resampling algorithm of Gordon, Salmond, and Smith (1993), the construction of the date-$t$ filter from the date-$(t-1)$ particles proceeds in two steps. In the mutation phase, a new set of particles is obtained by drawing a hidden state $\tilde{s}_t^{(n)}$ from each date-$(t-1)$ particle $s_{t-1}^{(n)}$ under the transition kernel of the Markov state. Given a new observation $r_t$, the particles are then resampled using weights that are proportional to the observation density $f_R(r_{t}|\tilde{s}_{t}^{(n)},R_{t-1}).$ Important refinements of the algorithm include sampling from an auxiliary model in the mutation phase (Pitt and Shephard, 1999), or implementing variance-reduction techniques such as stratified (Kitagawa 1996) and residual (Liu and Chen 1998) resampling. 

A common feature of existing filters is the requirement that the observation density $f_R(r_t|s_t,R_{t-1})$ be available analytically up to a normalizing constant. This condition need not hold in economic models in which equilibrium conditions can create complex nonlinear relationships between observations and the underlying state of the economy. In the special case when the state $s_t$ evolves in a Euclidean space $\mathbb{R}^{n_S}$ and has a continuous distribution, a possible solution is to estimate each observation density $f_R(r_{t}|\tilde{s}_{t}^{(n)},R_{t-1})$, $n\in\{1,\dots,N\}$, by nonparametric methods (Rossi and Vila, 2006, 2009). This approach is numerically challenging because $N$ conditional densities, and therefore $2 N^2$ kernels, must be evaluated every period. Furthermore, the rate of convergence decreases both with the dimension of the state space, $n_S$, and the dimension of the observation space, $n_R$, which indicates that the algorithm is prone to the curse of dimensionality.

The present paper develops a novel particle filter for general state space models that does not require the calculation of the observation density. This new method, which we call the State-Observation Sampling (SOS) filter, consists of simulating a state and a pseudo-observation $(\tilde{s}_t^{(n)},\tilde{r}_t^{(n)})$ from each date-$(t-1)$ particle. In the resampling stage, we assign to each particle $\tilde{s}_t^{(n)}$ an importance weight determined by the proximity between the pseudo-observation $\tilde{r}_{t}^{(n)}$ and the actual observation $r_{t}.$  We quantify proximity by a kernel of the type considered in nonparametric statistics: 
\begin{equation*}
p_t^{(n)}\propto \frac{1}{h_t^{n_R}}K\left(\frac{r_t-\tilde{r}_t^{(n)}}{h_t}\right),
\end{equation*}
where $h_t$ is a bandwidth, and $K$ is a probability density function. The resampling stage tends to select states associated with pseudo-observations in the neighborhood of the actual data. SOS requires the calculation of only $N$ kernels each period and makes no assumptions on the characteristics of the state space, which may or may not be Euclidean. We demonstrate that as the number of particles $N$ goes to infinity, the filter converges to the target distribution under a wide range of conditions on the bandwidth $h_t$. The root mean squared error of moments computed using the filter decays at the rate $N^{-2/(n_R+4)}$, that is at the same rate as the kernel density estimator of a random vector on $\mathbb{R}^{n_R}$. The asymptotic rate of convergence is thus invariant to the size of the state space, indicating that SOS overcomes a form of the curse of dimensionality. We also prove that the SOS filter provides consistent estimates of the likelihood function.

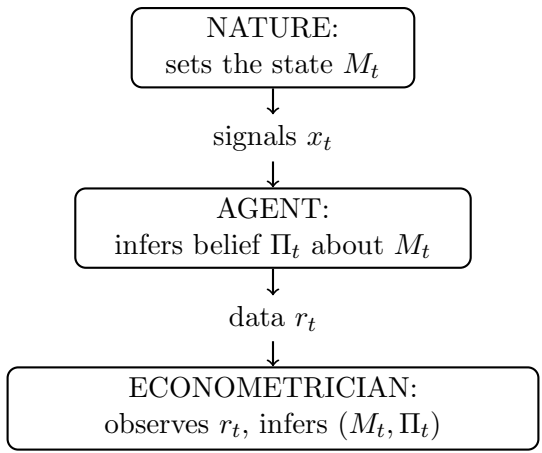
\begin{figure}
\begin{center}
\begin{tikzpicture}[edge from parent,
                     level distance=1.2cm,
                     grow=down,
                     parent anchor=south, child anchor=north]
  \tikzstyle{every node}=[rounded corners,text width=7.0cm, text badly centered,thick,black]
  \tikzstyle{edge from parent}=[draw,->,thick,black]       
  \node[draw,text width=3.5cm] at(0,0) {{\normalsize{NATURE:\\
        sets the state $M_t$}}}
   child{ node[fill=white]{signals $x_t$}
     child {node[draw, text width=5cm] {{\normalsize{AGENT:\\
  infers belief $\Pi_t$ about $M_t$}}}
     child{ node[fill=white]{data $r_t$}
       child {node[draw, text width=6.8cm] {\normalsize{ECONOMETRICIAN:\\
  observes $r_t$, infers $(M_t,\Pi_t)$}
                 } } } }}
   ;
 \end{tikzpicture}
\end{center}
\caption[]{Information structure.}
\label{Information Structure}
\end{figure}

We next develop inference methods for incomplete-information equilibrium models. 
To clarify the exposition, we focus on a class of recursive incomplete-information economies parameterized by $\theta\in\Theta,$ which nests the examples of Brandt, Zeng, and Zhang (2004), Calvet and Fisher (2007), David and Veronesi (2006), Lettau, Ludvigson and Wachter (2008), Moore and Schaller (1996) and van Nieuwerburgh and Veldkamp (2006). We consider three levels of information, which correspond to nature, an agent and the econometrician. Figure~\ref{Information Structure} illustrates the information structure. At the beginning of every period $t$, nature selects a Markov \textit{state of nature} $M_t$ and a vector of fundamentals or signals $x_t$, whose distribution is contingent on the state of nature. The agent observes the signal $x_t$, and computes the conditional probability distribution (``belief'') $\Pi_t=\Pi_t(x_t,\Pi_{t-1}),$ for instance by using Bayes' rule. According to her beliefs and signal, the agent also computes a data point $r_t=\mathcal{R}(x_t,\Pi_t,\Pi_{t-1};\theta),$ which may for example include asset returns, prices, or production decisions. The econometrician observes the data point $r_t$ and aims to track the hidden state $s_t=(M_t,\Pi_t)$ of the learning economy. 

We can apply the SOS filter to estimate the distribution of the state of the learning economy conditional on the observed data and the structural parameter $\theta$.
We propose an estimation procedure for $\theta$ based on indirect inference, a method introduced by Gouri\'{e}roux, Monfort and Renault (1993) and Smith (1993) that imputes the structural parameters of a model via an auxiliary estimator (e.g. Calzolari, Fiorentini and Sentana 2004; Czellar, Karolyi and Ronchetti 2007; Czellar and Ronchetti 2010; Dridi, Guay and Renault 2007; Genton and Ronchetti 2003; Heggland and Frigessi 2004). 
In our context, the full-information version of the economy, in which the state of nature $M_t$ is directly observed by the agent, is a natural building block of the auxiliary estimator. When the state of nature takes finitely many values, the Bayesian filter and the likelihood of the full-information model are available analytically (e.g. Hamilton, 1989). Similarly, when the state of nature $M_t$ has an infinite support, a full-information economy with discretized $M_t$ can be used. Given these properties, we define the auxiliary estimator by expanding the full-information economy's maximum likelihood estimator with a set of statistics that the incomplete-information model is designed to capture. 

We demonstrate the good performance of our techniques on a structural model of daily equity returns. Because the rich dynamics of the return series requires a large state space, we base our analysis on the multifrequency learning economy of Calvet and Fisher (``CF" 2007). We verify by Monte Carlo simulation that the SOS filter accurately tracks the state of the learning economy and provides remarkably precise estimates of the likelihood function. The indirect inference estimator is also shown to perform well in finite samples. 
We estimate the structural model on the daily excess returns of the CRSP U.S. value-weighted index between 1926 and 1999. For the out-of-sample period (2000-2009), the incomplete-information model provides accurate value-at-risk forecasts, which significantly outperform the predictions obtained from historical simulations, GARCH(1,1), and the full-information (FI) model.

The paper is organized as follows. Section 2 defines the SOS filter for general state space models. In section 3, we develop an indirect inference estimator for recursive learning economies. Section 4 applies these methods to a multifrequency investor learning model; we verify the accuracy of our inference methodology by Monte Carlo simulations, and conduct inference on the daily returns of a U.S. aggregate equity index between 1926 and 2009. Section 5 concludes.

\section{The State-Observation Sampling (SOS) Filter}\label{SOS}
\subsection{Definition}
We consider a discrete-time stochastic system defined on the probability space $(\Omega,\mathcal{F},\mathbb{P}).$ Time is discrete and indexed by $t=0,1,...,\infty.$ We consider a Markov process $s_t$ defined on a measurable space $(\mathcal{S},\mathcal{F}_S)$. For expositional simplicity, we assume in this subsection that $\mathcal{S}=\mathbb{R}^{n_S}$.

The econometrician receives every period an observation $r_t\in\mathbb{R}^{n_R}.$ Let $R_{t-1}=(r_1,...,r_{t-1})$ denote the vector of observations up to date $t-1$. The building block of our model is the conditional density of $(s_t,r_t)$ given $(s_{t-1},R_{t-1})$:
\begin{equation}
f_{S,R}(s_t,r_t|s_{t-1},R_{t-1}).
\label{stateobservation}
\end{equation}
Let $f_{S_0}$ denote a prior over the state space. The inference problem consists of estimating the density of the latent state $s_t$ conditional on the set of current and past observations: 
\begin{equation*}
f_{S}(s_t|R_t)
\end{equation*}
at all $t\geq1.$ 

A large literature proposes estimation by way of a particle filter, that is a finite set of points $(s_t^{(1)},...,s_t^{(N)})$ that targets $f_S(s_t|R_t)$. The sampling importance resampling method of Gordon, Salmond, and Smith (1993) is based on Bayes'rule:
\begin{equation*}
f_S(s_t|R_t)=\frac{f_R(r_t|s_t,R_{t-1})\,f_S(s_t|R_{t-1})}{f_R(r_t|R_{t-1})}\,.
\end{equation*}
The recursive construction begins by drawing $N$ independent states $s_0^{(1)},...,s_0^{(N)}$ from $f_{S_0}.$ Given the date$-(t-1)$ filter $(s_{t-1}^{(1)},\dots,s_{t-1}^{(N)})$, the construction of the date$-t$ filter proceeds in two steps. First, we sample $\tilde{s}_t^{(n)}$ from $s_{t-1}^{(n)}$ using the transition kernel of the Markov process. Second, in the resampling step, we sample $N$ particles $(s_t^{(1)},\dots,s_t^{(N)})$ from $(\tilde{s}_{t}^{(n)},\dots,\tilde{s}_{t}^{(N)})$ with normalized importance weights  
\begin{equation}
p^{(n)}_t = \frac{f_R(r_t|\tilde{s}^{(n)}_t,R_{t-1})}{\sum_{n'=1}^N f_R(r_t|\tilde{s}^{(n')}_t,R_{t-1})}.
\end{equation}
Under a wide range of conditions, the sample mean $N^{-1}\sum_{n=1}^N\Phi(s^{(n)}_t)$ converges to $\mathbb{E}[\Phi(s_t)|R_t]$ for any bounded measurable function $\Phi$.\footnote{See Crisan and Doucet (2002) for an excellent survey on the convergence of particle filters.}  

The sampling and importance resampling algorithm, and its various refinements, assume that the observation density $f_R(r_t|s_t,R_{t-1})$ is readily available up to a normalizing constant. This is a restrictive assumption in many applications, such as the incomplete-information economies considered in later sections. 

We propose a solution to this difficulty when it is possible to simulate from (\ref{stateobservation}). Our filter makes no assumption on the tractability of $f_{S,R}(\cdot|s_{t-1},R_{t-1})$, and in fact does not even require that the transitional kernel of the Markov state $s_t$ be available explicitly. The principle of our new filter is to simulate from each $s_{t-1}^{(n)}$ a state-observation pair $(\tilde{s}^{(n)}_t,\tilde{r}_t^{(n)})$, and then select particles $\tilde{s}_t^{(n)}$ associated with pseudo-observations $\tilde{r}_t^{(n)}$ that are close to the actual data point $r_t$. The definition of the importance weights is based on Bayes' rule applied to the joint distribution of $\tilde{r}^{(n)}_t$,\,$\tilde{s}^{(n)}_t$,\,$s_{t-1}^{(n)}$ conditional on $R_t$:
\begin{equation}
\tilde{r}^{(n)}_t,\tilde{s}^{(n)}_t,s_{t-1}^{(n)}|R_t \,\sim\, \frac{\delta(r_t-\tilde{r}^{(n)}_t) \, f_{S,R}(\tilde{s}^{(n)}_t,\tilde{r}^{(n)}_t|s_{t-1}^{(n)},R_{t-1})
\, f_S(s_{t-1}^{(n)}|R_{t-1})}{f_R(r_t|R_{t-1})},
\label{conditional density}
\end{equation}
where $\delta$ denotes the Dirac distribution on $\mathbb{R}^{n_R}$. Since the Dirac distribution produces degenerate weights, we consider a kernel $K$ with the following properties. 

\vskip10pt
\begin{assumption}\noindent\textbf{\upshape (Kernel). }\textit{The function $K:\mathbb{R}^{{n_R}}\rightarrow\mathbb{R}_{++}$ satisfies:
\begin{itemize}
\item[(i)]   $\int{K(u)du}=1$;
\item[(ii)]  $\int{uK(u)du}=0$; 
\item[(iii)] $A(K)=\int{\|u\|^2K(u)du} < \infty$;
\item[(iv)]  $B(K) = \int{[K(u)]^2du} < \infty.$
\label{kernel}
\end{itemize}
}
\end{assumption}
\vskip10pt

For any $r\in\mathbb{R}^{n_R},$ let 
\begin{equation*}
K_{h_t}(r)=\frac{1}{h_t^{n_R}}K\left(\frac{r}{h_t}\right)
\end{equation*} 
denote the corresponding kernel with bandwidth $h_t$ at date $t$. The kernel $K_{h_t}$ converges to the Dirac distribution as $h_t$ goes to zero, which we use to approximate (\ref{conditional density}). This suggests the following algorithm. 

\vskip15pt
\titledframe{\bf{SOS filter}}{
\parbox{13cm}{\underline{Step 1 (State-observation sampling)}: For every $n=1,\dots,N,$ we simulate a state-observation pair $(\tilde{s}_t^{(n)},\tilde{r}_t^{(n)})$ from  $f_{S,R}(\cdot|s_{t-1}^{(n)},R_{t-1})$.

\vskip10pt
\noindent
\underline{Step 2 (Importance weights)}: We observe the new data point $r_t$ and compute
$$p^{(n)}_t = \frac{K_{h_t} \left( r_t-\tilde{r}_t^{(n)}\right)}
{\sum_{n'=1}^N K_{h_t}\left(r_t - \tilde{r}_t^{(n')}\right)}, \text{  }n=1,\dots,N.$$

\vskip15pt
\noindent
\underline{Step 3 (Multinomial resampling)}: For every $n=1,\dots,N,$ we draw $s_t^{(n)}$ from $\tilde{s}_t^{(1)},\dots,\tilde{s}_t^{(N)}$ with importance weights $p^{(1)}_t,\dots,p^{(N)}_t$.}}
\vskip20pt

The state-observation pairs $\{(\tilde{s}_t^{(n)},\tilde{r}_t^{(n)})\}_{n=1,\dots,N}$ constructed in step 1 provide a discrete approximation to the conditional distribution of $(s_t,r_t)$ given the data $R_{t-1}$. In step 2, we construct a measure of the proximity between the pseudo and the actual data points, and in Step 3 we select particles for which this measure is large. The variance of multinomial resampling in step 3 can be reduced and computational speed can be improved by alternatives such as residual (Liu and Chen, 1998) or stratified (Kitagawa, 1996) resampling. In section 4, we obtain good results with a combined residual-stratified approach.\footnote{We select $\sum_{n=1}^N \lfloor N p^{(n)}_t \rfloor$ particles deterministically by setting $\lfloor N p^{(n)}_t \rfloor$ particles equal to $\tilde{s}_t^{(n)}$ for every $n\in\{1,\dots,N\},$ where $\lfloor\cdot\rfloor$ denotes the floor of a real number. 
The remaining $N_{r,t} = N - \sum_{n=1}^N \lfloor N  p^{(n)}_t \rfloor$ particles are selected by the stratified sampling that produces $\tilde{s}_t^{(n)}$ with probability $q^{(n)}_t = (N p^{(n)}_t - \lfloor N p^{(n)}_t \rfloor)/N_{r,t},$ $n=1,\dots,N$. That is, for every $k\in\{1,\dots,N_{r,t}\},$ we draw $\tilde{U}_k$ from the uniform distribution on $(\frac{k-1}{N_{r,t}},\frac{k}{N_{r,t}}]$, and select the particle $\tilde{s}_t^{(n)}$ such that $\tilde{U}_k\in(\sum_{j=1}^{n-1}q_t^{(j)},\sum_{j=1}^n q_t^{(j)}]$.} 
The convergence proof below applies equally well to these alternatives.

\subsection{Extension and Convergence}
The SOS filter easily extends to the case of a general measurable state space $\mathcal{S}.$ The building blocks of the model are the conditional probability measure of $(s_t,r_t)$ given $(s_{t-1},R_{t-1})$:
\begin{equation*}
g(\cdot|s_{t-1},R_{t-1}),
\end{equation*}
and a prior measure $\lambda_0$ over the state space. The SOS filter targets the probability measure of the latent state $s_t$ conditional on the set of current and past observations, $\lambda(\cdot|R_t).$
The SOS filter is defined as in Section~\ref{SOS}, where in step 1 we sample $(\tilde{s}_t^{(n)},\tilde{r}_t^{(n)})$ from the conditional measure $g(\cdot|s_{t-1}^{(n)},R_{t-1}).$ 

We now specify conditions under which for an arbitrary state space $\mathcal{S}$ and a fixed history $R_T=(r_1,\dots,r_T)$, $T\leq\infty$, the SOS filter converges in mean squared error to the target $\lambda(\cdot|R_t)$ as the number of particles $N$ goes to infinity.

\vskip10pt
\begin{assumption}\noindent\textbf{\upshape (Conditional Distributions). }\textit{The observation process satisfies the following hypotheses:
\begin{itemize}
\item[(i)] the conditional density $f_R(\tilde{r}_t|s_{t-1},R_{t-1})$ exists and \begin{equation*}
\kappa_t = \sup\{f_R(\tilde{r}_t|s_{t-1},R_{t-1});(s_{t-1},\tilde{r}_t)\in\mathcal{S}\times\mathbb{R}^{n_R}\} < \infty\,;
\end{equation*}
\item[(ii)] the observation density $f_R(\tilde{r}_t|s_t,R_{t-1})$ is well-defined and there exists $\kappa_t'\in\mathbb{R}_+$ such that:
\begin{equation*}
|f_R(\tilde{r}_t|s_t,R_{t-1})-f_R(r_t|s_t,R_{t-1})-\frac{\partial f_R}{\partial r_t'}(r_t|s_t,R_{t-1})(\tilde{r}_t-r_t)|\leq \kappa_t' \|\tilde{r}_t-r_t\|^{2}
\end{equation*}
for all $(s_t,\tilde{r}_t) \in \mathcal{S}\times\mathbb{R}^{n_R}$ and $t\leq T$.
\end{itemize}}
\label{f2bound} 
\end{assumption}
\vskip10pt

\vskip10pt
\begin{assumption}\noindent\textbf{\upshape (Bandwidth).}\textit{
The bandwidth is a function of $N$, $h_t=h_t(N)$, and satisfies
\begin{itemize}
\item[(i)] $\lim_{N\rightarrow\infty}h_t(N) = 0,$ 
\item[(ii)] $\lim_{N\rightarrow\infty}N[h_t(N)]^{n_R} = +\infty,$ 
\end{itemize}
for all $t=1,\dots,T.$}
\label{bandwidth}
\end{assumption}
\vskip10pt
We establish the following result in the appendix.
\vskip10pt

\begin{thm}\noindent\textbf{\upshape (Convergence of the SOS Filter). }\textit{ 
Under assumptions~\ref{kernel} and \ref{f2bound} and for every $t$ and $N\geq1$, there exists $U_t(N)\in\mathbb{R_+}$ such that
\begin{equation}
\mathbb{E}\left\{\left[\frac{1}{N}\sum_{n=1}^N K_{h_t}(r_t - \tilde{r}_t^{(n)})-f_R(r_t|R_{t-1})\right]^2\right\}\leq \frac{[f_R(r_t|R_{t-1})]^2}{4} U_t(N), 
\label{likelihood_bound}
\end{equation}
where the expectation is over all the realizations of the random particle method. Furthermore, for any bounded measurable function, $\Phi:\mathcal{S}\rightarrow\mathbb{R},$
\begin{equation} 
MSE_t = \mathbb{E}\left\{\left[\frac{1}{N}\sum_{n=1}^N \Phi(s_{t}^{(n)}) - \mathbb{E}[\Phi(s_t)|R_t]\right]^2\right\} \leq U_t(N)\|\Phi\|^2,
\label{MSEt}
\end{equation}
where $\|\Phi\|=\sup_{s\in\mathcal{S}}|\Phi(s)|.$
If assumption~\ref{bandwidth} also holds, then 
\begin{equation*}
\lim_{N\to\infty}U_t(N)=0\,,
\end{equation*}
and the filter converges in mean squared error. Furthermore, if the bandwidth sequence is of the form $h_t(N)=h_t(1)N^{-1/(n_R+4)}$, then $U_t(N)$ decays at rate $N^{-4/(n_R+4)}$ and the root mean squared error $MSE_t^{1/2}$ at rate $N^{-2/(n_R+4)}$ for all $t.$
}
\label{theo}
\end{thm}

\vskip10pt

By (\ref{likelihood_bound}), the kernel estimator 
\begin{equation}
\hat{f}_R(r_t|R_{t-1}) = \frac{1}{N} \sum_{n=1}^N K_{h_t}(r_t-\tilde{r}_t^{(n)}),
\end{equation}
converges to the conditional density of $r_t$ given past observations. Consequently, we can estimate the log-likelihood function by $\sum_{t=1}^T \ln\hat{f}_R(r_t|R_{t-1}),$ and provide a plug-in bandwidth in the online Appendix. We will illustrate in section 4 the finite-sample accuracy of the SOS filter.

\section{Recursive Learning Economies}\label{learning}
We consider a class of discrete-time stochastic economies defined at $t=0,\dots,\infty$ on the probability space $(\Omega,\mathcal{F},\mathbb{P})$ and parameterized by $\theta\in\Theta\subseteq\mathbb{R}^p,$ $p\geq1.$ 

\subsection{Information Structure}
In every period $t,$ we define three levels of information, which respectively correspond to nature, a Bayesian agent, and the econometrician. Figure~\ref{Information Structure} illustrates the information structure.

\subsubsection{Nature}
A \textit{state of nature} $M_t$ drives the fundamentals of the economy. We assume that $M_t$ follows a first-order Markov chain on the set of mutually distinct states $\{m^1(\theta),\dots,m^d(\theta)\}.$ For every $i,j\in\{1,..,d\},$ we denote by $a_{i,j}(\theta) = \mathbb{P}(M_{t}=m^j(\theta)|M_{t-1}=m^i(\theta);\theta)$ the transition probability from state $i$ to state $j.$ We assume that the Markov chain $M_t$ is irreducible, aperiodic, positive recurrent, and therefore ergodic. For notational simplicity, we henceforth drop the argument $\theta$ from the states $m^j$ and transition probabilities $a_{i,j}.$ 

\subsubsection{Agent}
At the beginning of every period $t$, the agent observes a signal vector $x_t\in \mathbb{R}^{n_X}$, which is partially revealing on the state of nature $M_t$. The probability density function of the signal conditional on the state of nature, $f_X(x_t|M_t;\theta),$ is known to the agent. Let $X_t=(x_1,\dots,x_t)$ denote the vector of signals received by the agent up to date $t$. For tractability reasons, we make the following hypotheses.

\vskip10pt
\noindent
\begin{assumption} \textbf{\upshape (Signal). }\textit{The signal satisfies the following conditions:
\begin{itemize}
\item[(a)] $\mathbb{P}(M_t=m^j|M_{t-1}=m^i,X_{t-1};\theta)=a_{i,j}$ for all $i,j$ ; 
\item[(b)] $f_X(x_t|M_t,M_{t-1},\dots,M_0,X_{t-1};\theta)=f_X(x_t|M_t;\theta).$ 
\end{itemize}}
\label{signal}
\end{assumption}
\vskip10pt

\noindent The agent knows the structural parameter $\theta,$ is Bayesian and uses $X_t$ to compute the conditional probability of the states of nature.

\vskip10pt
\begin{prop} \upshape{\bf(Agent Belief).} \it{ The conditional probabilities 
$\Pi_t^j = \mathbb{P}(M_t=m^j|X_t;\theta)$ satisfy the recursion:
\begin{equation}
\Pi_t^{j} = \frac{\omega^j(\Pi_{t-1},x_t;\theta)}{\sum_{i=1}^d \omega^i(\Pi_{t-1},x_t;\theta) }\text{   for all $j\in\{1,\dots,d\}$ and $t \geq 1,$}
\label{Bayes}
\end{equation}
where $\Pi_{t-1}= (\Pi_{t-1}^1,\dots,\Pi_{t-1}^d)$ and $\omega^j(\Pi_{t-1},x_t;\theta) = f_X(x_t|M_t=m^j;\theta) \sum_{i=1}^d a_{i,j} \Pi_{t-1}^{i}.$}
\label{AgentBelief}
\end{prop}
\vskip10pt
\noindent In applications, the agent values assets or makes financial, production or purchasing decisions as a function of the belief vector $\Pi_t.$ Our methodology easily extends to learning models with non-Bayesian agents, as in Brandt, Zeng, and Zhang (2004) and Cecchetti Lam and Mark (2000).

The \textit{state of the learning economy} at a given date $t$ is the mixed variable $s_t = (M_t,\Pi_t)$. The state space is therefore
\begin{equation}
\mathcal{S}=\{m^1,\dots,m^d\}\times\Delta_{+}^{d-1},
\label{statespace}
\end{equation}
where $\Delta_{+}^{d-1}=\{\Pi\in\mathbb{R}_+^d|\sum_{i=1}^d \Pi_i=1\}$ denotes the $(d-1)$--dimensional unit simplex. 

\vskip10pt
\begin{prop} \noindent\textbf{\upshape (State of the Learning Economy). }\textit{The state of the learning economy, $s_t=(M_t,\Pi_t),$ is first-order Markov. It is ergodic if the transition probabilities between states of nature are strictly positive: $a_{i,j}>0$ for all $i,j$, and the signal's conditional probability density functions $f_X(x|M_t=m^j;\theta)$ are strictly positive for all $x\in\mathbb{R}^{n_X}$ and $j\in\{1,\dots,d\}$.}
\label{ergodic}
\end{prop} 
\vskip10pt

\noindent The state of the learning economy $s_t$ preserves the first-order Markov structure of the state of nature $M_t.$ By Bayes'rule (\ref{Bayes}), the transition kernel of the Markov state $s_t$ is sparse when the dimension of the signal, $n_X,$ is lower than the number of states of nature: $n_X<d$. The state $s_t$ is nonetheless ergodic for all values $n_X$ and $d$ under the conditions stated in Proposition~\ref{ergodic}, which guarantees that the economy is asymptotically independent of the initial state $s_0.$

\subsubsection{Econometrician}
Each period, the econometrician observes a data point $r_t\in\mathbb{R}^{n_R},$ which is assumed to be a deterministic function of the agent's signal and conditional probabilities over states of nature: 
\begin{equation}
r_t=\mathcal{R}(x_t,\Pi_t,\Pi_{t-1};\theta).
\label{return_def}
\end{equation}
We include $\Pi_{t-1}$ in this definition to accommodate the possibility that $r_t$ is a growth rate or return. The parameter vector $\theta\in\mathbb{R}^p$ specifies the states of nature $m^1,\dots,m^d,$ their transition probabilities $(a_{i,j})_{1\leq i,j \leq d},$ the signal's conditional density $f_X (\cdot|M_t,\theta),$ and the data function $\mathcal{R}(x_t,\Pi_t,\Pi_{t-1};\theta).$ In some applications, it may be useful to add measurement error in (\ref{return_def}); the estimation procedure of the next section applies equally well to this extension.

\subsection{Estimation}\label{IIsec}
We assume that the data $R_T=(r_1,\dots,r_T)$ is generated by the incomplete-information (II) economy with parameter $\theta^*$ described above. Estimation faces several challenges. The transition kernel of the Markov state $s_t$ and the log-likelihood function $\mathcal{L}_{II}(\theta|R_T)$ are not available analytically. Furthermore, the observation density $f_R(r_t|s_t,R_{t-1})$ is not available in closed form either because the signal $x_t,$ drives the data point $r_t=\mathcal{R}(x_t,\Pi_t,\Pi_{t-1};\theta)$ both directly and indirectly through the belief $\Pi_t=\Pi_t(x_t,\Pi_{t-1}),$ creating a highly nonlinear relationship between the state and the observation.  

The learning model can, however, be conveniently simulated. Given a state $s_{t-1}=(M_{t-1},\Pi_{t-1})$, we can: $(i)$ sample $M_{t}$ from $M_{t-1}$ using the transition probabilities $a_{i,j}$; $(ii)$ sample the signal $x_{t}$ from $f_X(\cdot|M_{t};\theta)$; $(iii)$ apply Bayes'rule (\ref{Bayes}) to impute the agent's belief $\Pi_{t}$; and $(iv)$ compute the simulated data point $\tilde{r}_{t}=\mathcal{R}(x_t,\Pi_t,\Pi_{t-1};\theta)$. Estimation can therefore proceed by simulation-based methods. Simulated ML based on the SOS filter is a possible approach. As we will see in section 4, however, an accurate approximation of the log-likelihood value  $\hat{\mathcal{L}}_{II}(\theta|R_T)=\sum_{t=1}^T{\text{ln}}[N^{-1}\sum_{n=1}^N K_{h_t}(r_t-\tilde{r}_t^{(n)})]$ may require a large number of particles. For situations where simulated ML is too computational\footnote{For instance in the empirical example considered in section 4, we use an SOS filter of size $N=10^7$ and a dataset of about 20,000 observations. One evaluation of the likelihood function requires the evaluation 200 billion kernels $K_{h_t}(\cdot)$. Since a typical optimization requires about 500 function evaluations, the simulated ML estimation of the II model would require the evaluation of 100 trillion kernels.}, 
we now propose an alternative approach based on indirect inference. 

For each learning model $\theta\in\Theta$, we can define an auxiliary {\it full information} (FI) model in which the agent observes both the state of nature $M_t$ and the signal $x_t.$ Her conditional probabilities are then  $\Pi_t^j=\mathbb{P}(M_t=m^j | X_t, M_t;\theta)$ for all $j$. The belief vector reduces to $\Pi_t={\mathds{1}}_{M_t}$, where ${\mathds{1}}_{M_t}$ denotes the vector whose $j^{th}$ component is equal to 
1 if $M_t=m^j$ and 0 otherwise, and by (\ref{return_def}) the full information data point is defined by $r_t=\mathcal{R}(x_t,\mathds{1}_{M_t},\mathds{1}_{M_{t-1}};\theta)$. The FI model can have less parameters than the II model because of the simplification in $\Pi_t$. We therefore consider that the auxiliary FI model is parameterized by $\phi\in\mathbb{R}^q$, where $1\leq q\leq p.$ 

\vskip10pt
\begin{assumption}
\textbf{\upshape{ (Auxiliary Full-Information Economies). }}\textit{The probability density functions $f_{i,j}(r_t;\phi)=f_{R,FI}(r_t|M_t=m^j,M_{t-1}=m^i,\phi)$ are available analytically for all $i,j\in\{1,\dots,d\}$.}
\label{closefR}
\end{assumption}

\vskip10pt
\begin{prop} \upshape{\bf(Full-Information Likelihood).} \it{Under assumption \ref{closefR}, the log-likelihood function $\mathcal{L}_{FI}(\phi|R_{T})$ is available analytically.}
\label{closedformL}
\end{prop}
\vskip10pt

\noindent The ML estimator of the full-information economy
\begin{equation*}
\hat{\phi}_T=\arg\max_{\phi}\mathcal{L}_{FI}(\phi|R_T)\in\mathbb{R}^q
\end{equation*} 
can therefore be conveniently computed.

The indirect-inference estimation of the structural learning model proceeds in two steps. First, we define an auxiliary estimator that includes the full-information MLE. If $q<p,$ we also consider a set of $p-q$ statistics $\hat\eta_T$ that quantify features of the dataset $R_T$ that the learning model is designed to capture. The \textit{auxiliary estimator} is defined by
\begin{equation}
\label{auxest}
   \hat{\mu}_T= \left[\begin{array}{l}
												\hat{\phi}_T\\
												\hat{\eta}_T 
										 \end{array} \right] \in \mathbb{R}^p.
\end{equation}
By construction, $\hat{\mu}_T$ contains as many parameters as the structural parameter $\theta.$\footnote{We focus on the exactly identified case to simplify the exposition and because earlier evidence indicates that parsimonious auxiliary models tend to provide more accurate inference in finite samples (e.g. Andersen, Chung, and Sorensen, 1999; Czellar and Ronchetti, 2010). Our approach naturally extends to the overidentified case, which may be useful in cases where it is economically important to match a larger set of statistics.}

Second, for any admissible parameter $\theta$, we can simulate a sample path $R_{ST}(\theta)$ of length $ST$, $S\geq 1,$ and compute the corresponding pseudo-auxiliary estimator:
\begin{equation}
\hat\mu_{ST}(\theta)=  \left[\begin{array}{l}
												\hat\phi_{ST}(\theta)\\
												\hat{\eta}_{ST}(\theta) 
										 \end{array} \right],
\end{equation}
where $\hat\phi_{ST}(\theta)=\arg\max_{\phi}\mathcal{L}_{FI}[\phi|R_{ST}(\theta)]$. We define the indirect inference estimator $\hat\theta_{T}$ by: 
\begin{equation}
\hat\theta_{T}=\arg\min_\theta\big[\hat\mu_{ST}(\theta)-\hat\mu_T\big]'\Omega\big[\hat\mu_{ST}(\theta)-\hat\mu_T\big]\,,
\label{II}
\end{equation}
where $\Omega$ is a positive definite weighting matrix. When the calculation of the full-information MLE is expensive, the numerical implementation can be accelerated by the efficient method of moments, as is discussed in the appendix. 

Our methodology builds on the fact that the full-information economy can be efficiently estimated by ML and is therefore a natural candidate auxiliary model. Moreover, the theoretical investigation of a learning model often begins with the characterization of the FI case, so the estimation method we are proposing follows the natural progression commonly used in the literature.

We assume that the assumptions \ref{binding function}--\ref{Hessian} given in the appendix hold. Gouri\'eroux et al. (1993) and Gouri\'eroux and Monfort (1996) show that under these conditions and assuming the structural model $\theta^*$, the auxiliary estimator $\hat{\mu}_T$ converges in probability to a deterministic function $\mu(\theta^*)$, called the \textit{binding function}, and $\sqrt{T}\left[\hat{\mu}_T-\mu(\theta^*)\right]\overset{d}{\longrightarrow} \mathcal{N}(0,W^*),$ where $W^*$ is defined in the appendix. Furthermore, when $S$ is fixed and $T$ goes to infinity, the estimator $\hat\theta_{T}$ is consistent and asymptotically normal:
\begin{equation*}
\sqrt{T}(\hat\theta_{T}-\theta^*)\overset{d}{\longrightarrow} \mathcal{N}(0,\Sigma),
\end{equation*}
where
\begin{equation} 
  \Sigma = 
  \left(1+\frac{1}{S}\right)
	\left[\frac{\partial \mu(\theta^*)}{\partial\theta'}\right]^{-1}	W^*   	
	\left[\frac{\partial \mu(\theta^*)'}{\partial\theta}
	\right]^{-1}\,.
	\label{IIvar}
\end{equation}
\noindent The appendix further discusses the numerical implementation of this method.

In this section, we have assumed that the state of nature takes finitely many values. When $M_t$ has an infinite support, we can discretize its distribution and use the corresponding full-information discretized economy as an auxiliary model. The definition and properties of the indirect inference estimator are otherwise identical.

\section{Inference in an Asset Pricing Model with Investor Learning}
We now apply our methodology to a consumption-based asset pricing model. We adopt the Lucas tree economy with regime-switching fundamentals of CF (2007), which we use to specify the dynamics of daily equity returns.

\subsection{Specification}
\subsubsection{Dynamics of the State of Nature}
The rich dynamics of daily returns requires a large state space. For this reason, we consider that the state is a vector containing $\overline{k}$ components:
\begin{equation*}
M_t=(M_{1,t},\dots,M_{\overline{k},t})'\in\mathbb{R}_+^{\overline{k}},
\end{equation*}
which follows a binomial Markov Switching Multifractal (CF 2001, 2004, 2008). The components are mutually independent across $k.$ Let $M$ denote a Bernoulli distribution that takes either a high value $m_0$ or a low value $2-m_0$ with equal probability. Given a value $M_{k,t}$ for the $k^{th}$ component at date $t$, the next-period multiplier $M_{k,t+1}$ is either:

\begin{equation*}
\begin{cases}
{\text{drawn from the distribution $M$ with  probability $\gamma_k$},}\\
{\text{equal to its current value $M_{k,t}$ with  probability $1-\gamma_k$.}}
\end{cases}
\end{equation*}
Since each component of the state vector can take two possible values, the state space contains $d=2^{\overline{k}}$ elements $m^1,\dots,m^d$. The transition probabilities $\gamma_k$ are parameterized by
\begin{equation*}
\gamma_k=1-(1-\gamma_{\overline{k}})^{b^{k-\overline{k}}},\text{  } k=1,\dots,\overline{k},
\end{equation*}
where $b>1.$ Thus, $\gamma_{\overline{k}}$ controls the persistence of the highest-frequency component and $b$ determines the spacing between frequencies.

\subsubsection{Bayesian Agent} 
The agent receives an exogenous consumption stream $\{C_t\}$ and prices the stock, which is a claim on an exogenous dividend stream $\{D_t\}$. Every period, the agent observes a signal $x_t\in{\mathbb{R}}^{\overline{k}+2}$ consisting of dividend growth:
\begin{equation}
x_{1,t}=\ln(D_t/D_{t-1})=g_D-\frac{\sigma^2_D(M_t)}{2}+\sigma_D(M_t)\varepsilon_{D,t},
\label{exogenous dividend}
\end{equation}
consumption growth:
\begin{equation}
x_{2,t}=\ln(C_t/C_{t-1})=g_C+\sigma_C\varepsilon_{C,t},
\label{exogenous consumption}
\end{equation}
and a noisy version of the state:
\begin{equation}
x_{i+2,t}=M_{i,t}+\sigma_\delta z_{i,t},\hskip 10pt i=1,\dots,\overline{k}\,.
\end{equation}
The noise parameter $\sigma_\delta\in\mathbb{R}_+$ controls information quality. The stochastic volatility of dividends is given by:
\begin{equation}
\sigma_D(M_t)=\overline\sigma_D\left(\prod_{k=1}^{\overline{k}}M_{k,t}\right)^{1/2},
\end{equation}
where $\overline\sigma_D\in\mathbb{R}_+.$ The innovations $\varepsilon_{C,t},$ $\varepsilon_{D,t},$ and $z_t$ are jointly normal and have zero means and unit variances. We assume that $\varepsilon_{C,t}$ and $\varepsilon_{D,t}$ have correlation $\rho_{C,D},$ and that all the other correlation coefficients are zero.

Learning about the volatility state $M_t$ is an asymmetric process. For expositional simplicity, assume that the noise parameter $\sigma_\delta$ is large, so that investors learn about $M_t$ primarily through the dividend growth. Because large realizations of dividend are implausible in a low-volatility regime, learning about a volatility increase tends to be abrupt. Conversely, when volatility switches from a high to a low state, the agent learns only gradually that volatility has gone down because realizations of dividend growth near the mean are likely outcomes under any $M_t.$ 

The agent has isoelastic expected utility,
$U_0 = \mathbb{E}_0 \sum_{t=0}^\infty \delta^t C_t^{1-\alpha}/(1-\alpha),$
where $\delta$ is the discount rate and $\alpha$ is the coefficient of relative risk aversion. In equilibrium, the log interest rate is constant. The stock's price-dividend ratio is negatively related to volatility and linear in the belief vector: 
\begin{equation}
Q(\Pi_t)=\sum_{j=1}^d{Q(m^j) \Pi_t^{j}.}
\label{Qpi}
\end{equation}
where the linear coefficients $Q(m^j)$ are available analytically.\footnote{The price-dividend ratio is given by
$$\sum_{n=1}^{\infty}\delta^n \mathbb{E}\left[\left.\left(\frac{C_{t+n}}{C_t}\right)^{-\alpha} \frac{D_{t+n}}{D_t}\right|X_t\right] 
=\sum_{n=1}^{\infty}\mathbb{E}\left[\left.\prod_{h=1}^n e^{g_D-r_f-\alpha\rho_{C,D}\sigma_C\sigma_D(M_{t+h})}\right|X_t\right],$$
where $r_f=-\ln(\delta)+\alpha g_C-\alpha^2\sigma_C^2/2$ is the log interest rate. Since volatility is persistent, a high level of volatility at date $t$ implies high forecasts of future volatility, and therefore a low period$-t$ price-dividend ratio. The linear coefficients are given by $\big(Q(m^1),\dots,Q(m^d)\big)'=(I-B)^{-1}\iota - \iota\,,$
where $B=(b_{ij})_{1 \leq i,j \leq d}$ is the matrix with components $b_{ij}= a_{i,j} \exp\big[{g_D-r_f-\alpha\,\rho_{C,D}\,\sigma_C\,\sigma_D(m^j)}\big]$ and $\iota = (1,\dots,1)'$.}

\subsubsection{Econometric Specification of Stock Returns}

\begin{figure}
\includegraphics[height=0.5\textheight]{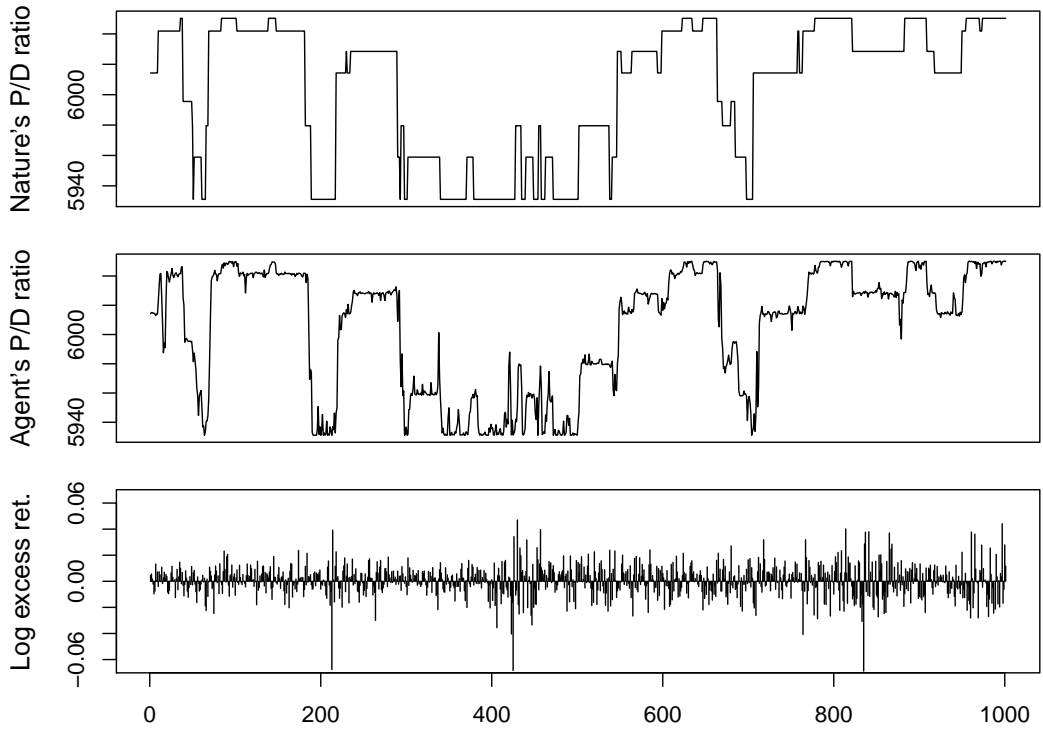}
\caption[]{Learning Model Simulation. This figure illustrates a sample path simulated from the multifrequency learning model. Each panel corresponds to a different level of information. Nature's price-dividend ratio $Q(M_t)$ is plotted in the top panel, the agent's price-dividend ratio $Q(\Pi_t)$ in the middle panel, and the return $r_t$ (computed by the agent and observed by the econometrician) in the bottom panel.} 
\label{rtQtrue}
\end{figure}

The econometrician observes the log excess return process:
\begin{equation}
r_t = \ln\left[\frac{1+Q(\Pi_{t})}{Q(\Pi_{t-1})}\right]+x_{1,t}-r_f\,.
\label{return}
\end{equation}
Since learning about volatility is asymmetric, the stock price falls abruptly following a volatility increase (bad news), but will increase only gradually after a volatility decrease (good news). The noise parameter $\sigma_{\delta}$ therefore controls the skewness of stock returns.

\subsection{Accuracy of the SOS filter}
We now present the results of Monte Carlo simulations of the SOS particle filter. To simplify the exposition, we consider one-dimensional aggregates of $M_t$ and $\Pi_t,$ which summarize economically meaningful information. Specifically, if the agent knew the true state of nature, she would set the price-dividend ratio equal to $Q(M_t)=Q(m^j)$ if  $M_t=m^j,$ as implied by (\ref{Qpi}); we therefore call $Q(M_t)$ {\it nature's $P/D$ ratio.} By contrast, the market $Q(\Pi_t)$ aggregates the agent's beliefs in the incomplete-information model; for this reason, we refer to it as the {\it agent's price-dividend ratio}. 

\begin{figure}
\includegraphics[height=0.45\textheight]{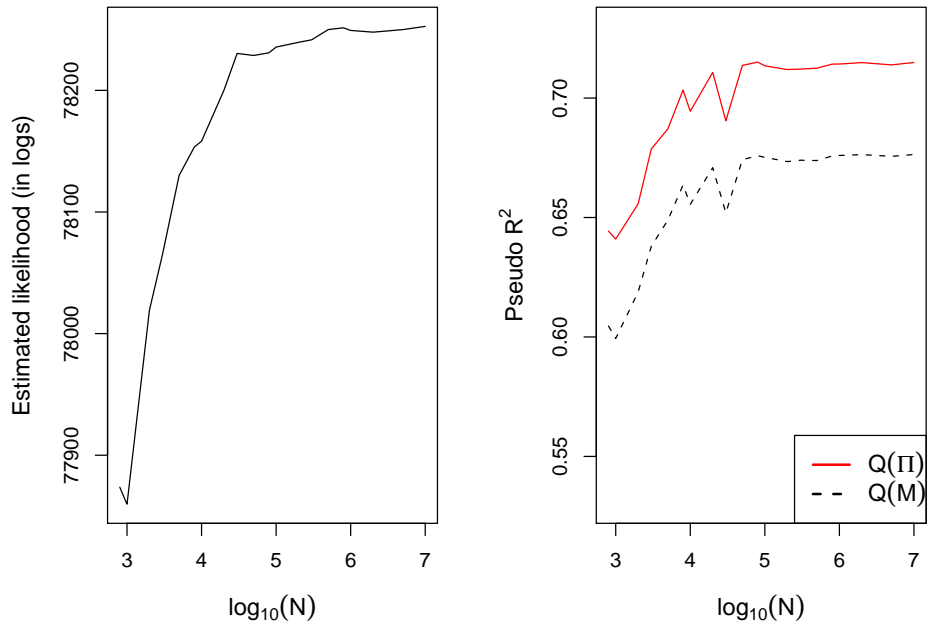}
\caption[]{Accuracy of the SOS Filter. This figure illustrates the estimated log-likelihood function (left panel) and the efficiency measures $R_{Q(\Pi)}^2$ and $R_{Q(M)}^2$ (right panel) as a function of the filter size $N$.} 
\label{Particles}
\end{figure}

We generate a sample of size $T=20,000$ periods from the learning model (\ref{return}) with $\overline{k}=3$ volatility components and fixed parameter values.\footnote{Specifically, we set $m_0=1.7$, $\gamma_{\overline{k}}=0.06$, $b=2$ and $\sigma_\delta=1$, the consumption drift to $g_C=0.75$ basis point (bp) (or 1.18\% per year), log interest rate to $r_f=0.42$ bp per day ($1\%$ per year), excess dividend growth equal to $g_D-r_f=0.5$ bp per day (about $1.2\%$ per year), consumption volatility to $\sigma_C=0.189\%$ (or 2.93\% per year), and dividend volatility $\overline{\sigma}_{D}=0.70\%$ per day (about $11\%$ per year). The correlation coefficient is set equal to $\rho_{C,D}=0.6$, and $\alpha$ is chosen such that the mean of the linear coefficients in (\ref{Qpi}) satisfy $\overline{Q}=d^{-1}\sum_{i=1}^dQ(m^i)=6000$ in daily units (25 in yearly units).}
Figure~\ref{rtQtrue} illustrates the last 1,000 periods of the simulated sample. 
We report nature's price-dividend ratio in the top panel, the agent's price-dividend ratio in the middle panel, and the return (computed by the agent and observed by the econometrician) in the bottom panel.

We apply to the entire simulated sample the SOS filter with the quasi-Cauchy kernel and bandwidth derived in the online Appendix. The left panel of Figure~\ref{Particles} illustrates the estimated log-likelihood as a function of the filter size $N.$ In the right panel, we report the pseudo $R^2$:
\begin{equation*}
R_{Q(\Pi)}^2= 1 - \frac{\sum_{t=1}^T \big[\hat{Q}(\Pi_t)-Q(\Pi_t)\big]^2}{\sum_{t=1}^T \big[\hat{Q}(\Pi_t) - \bar{Q}(\Pi)\big]^2}\,,
\end{equation*}
where $\hat{Q}(\Pi_t) = \sum_{n=1}^N Q(\Pi^{(n)}_t)/N$ and $\bar{Q}(\Pi) = \sum_{t=1}^T Q(\Pi_t) /T$. We similarly compute $R^2_{Q(M)}$ for nature's price-dividend ratios using $\{Q(M_t^{(n)})\}$. The figure shows that both the estimated log-likelihood and the coefficients of determination increase with the filter size $N$ and settle down for $N\geq 10^6$. The coefficient of determination reaches $67.6\%$ for $Q(M)$ and $71.5\%$ for $Q(\Pi)$. Thus, the agent's P/D ratio is better estimated than nature's P/D ratio, as the information structure in Figure~\ref{Information Structure} suggests.
  
The true value of the likelihood function is unknown for the example considered in Figure~\ref{Particles}. For this reason, we now consider the full-information version of the model, which, by Proposition~\ref{closedformL}, has a closed-form likelihood. We generate from the full-information model a sample of $T=20,000$ periods. The analytical expression of the log-likelihood implies that $\mathcal{L}_{FI}=79,691.5$. In the right column of Table~\ref{SOSramse}, we  report the sample mean and the root mean squared error of fifty log-likelihood estimates computed using SOS. The relative estimation error RMSE/$\mathcal{L}_{FI}$ is 0.024\%, 0.006\% and 0.002\% when using, respectively, $N=10^5$,\,$10^6$ and $10^7$ particles. The estimates of the FI log-likelihood obtained using SOS are therefore remarkably precise.

We now verify that the SOS filter defeats the curse of dimensionality with respect to the size of the state space. Table~\ref{IIFIns} reports the topological dimension of the state space, dim$\,\mathcal{S}$, under incomplete and full information. By construction, the log-likelihood function satisfies the continuity property: $\lim_{\sigma_\delta\to 0}\mathcal{L}_{II}(m_0,\gamma_{\overline{k}},b,\sigma_\delta|R_T)=\mathcal{L}_{FI}(m_0,\gamma_{\overline{k}},b|R_T)\,.
$
The first three columns in Table~\ref{SOSramse} report summary statistics of log-likelihood estimates of $\mathcal{L}_{II}$ obtained for $\sigma_\delta\in\{1,\,0.1,\,0.01\}$. The accuracy of SOS is nearly identical for the full-information model and for the learning model with $\sigma_\delta=0.01.$ With $N=10^7$ particles, the RMSE of the SOS filter is even slightly smaller for the II specification $\sigma_\delta=0.01$ than for the full-information model, even though II has a much larger state space. These findings confirm the result of Theorem~\ref{theo} that the convergence rate of SOS is independent of the dimension of the state space.
  
\begin{table}
\begin{center}
\begin{minipage}[t]{11.5cm}
   \renewcommand{\footnoterule}{} 
\caption{{\sc{Dimension of the state space}}\protect\footnote{This table reports the topological dimension of the state space under full and incomplete information. In the multifrequency volatility case, we know that $d=2^{\overline{k}}$, where $\bar{k}$ denote the number of volatility frequencies.
}}
\vskip5pt
\begin{tabular}{l|c|c}
\hline\hline
\tvi & Incomplete Information & Full Information\\
\hline
\tvi State space $\mathcal{S}$ & $\{m^1,\dots,m^d\}\times\Delta_{+}^{d-1}$ & $\{m^1,\dots,m^d\}$\\
\hline
\tvi Dimension dim$\,\mathcal{S}$ & $d-1$ & 0\\
\hline
\end{tabular}
\label{IIFIns}
\end{minipage}
\end{center}
\end{table}

\vskip5pt
\begin{table}
\begin{center}
\begin{minipage}[t]{12.7cm}
   \renewcommand{\footnoterule}{} 
\caption{{\sc{Precision of the SOS log-likelihood estimates}}\protect\footnote{We report summary statistics for 50 simulated log-likelihoods estimated on a fixed sample path of $T=20,000$ periods from the FI model. The true log-likelihood is $\mathcal{L}_{FI}=79,691.5$. The simulated log-likelihoods are based on an SOS filter and a learning model with $\sigma_\delta\in\{0.01,0.1,1\}$.}}
\vskip5pt
\begin{tabular}{l|rrr|r}
\hline\hline
\tvi & \multicolumn{3}{c}{II (dim\,$\mathcal{S}=7$)} &  \multicolumn{1}{|c}{FI (dim\,$\mathcal{S}=0$)}\\
\tvi & $\sigma_\delta=1$ & $\sigma_\delta=0.1$ & $\sigma_\delta=0.01$ & \\
\hline
\tvi
Mean, $N=10^5$ & 79,514.1 & 79,674.0 & 79,673.1 & 79,673.4\\
Mean, $N=10^6$ & 79,523.2 & 79,686.3 & 79,687.4 & 79,687.3\\
Mean, $N=10^7$ & 79,525.1 & 79,690.8 & 79,690.9 & 79,690.4\\
\hline
\tvi
RMSE, $N=10^5$ & 177.6 & 18.3 & 19.5 & 18.9\\
RMSE, $N=10^6$ & 168.4 & 6.3 & 5.2 & 4.9\\
RMSE, $N=10^7$ & 166.4 & 1.1 & 1.3 & 1.7\\
\hline
\end{tabular}
\label{SOSramse}
\end{minipage}
\end{center}
\end{table}

\subsection{Indirect Inference Estimator}
We now develop an estimator for the vector of structural parameters:
\begin{equation*}
\theta=(m_0,\gamma_{\overline{k}},b,\sigma_\delta)'\in [1,2]\times\left(0,1\right]\times[1,\infty)\times\mathbb{R}_+,
\end{equation*} 
where $m_0$ controls the variability of dividend volatility, $\gamma_{\bar{k}}$ the transition probability of the most transitory volatility component, $b$ the spacing of the transition probabilities, and $\sigma_\delta$ the precision of the signal received by the representative agent. 
As is traditional in the asset pricing literature, we calibrate all the other parameters on aggregate consumption data and constrain the mean price-dividend ratio to a plausible long-run value
\begin{equation}
\mathbb{E}[Q(\Pi_t)] = \overline{Q},
\label{Qbar}
\end{equation}
where $\overline{Q}$ is set equal to 25 in yearly units.\footnote{The calibrated parameters are the same as in the previous subsection. An alternative approach would be to estimate all the parameters of the learning economy on aggregate excess return data. In the 2005 NBER version of their paper, CF applied this method to the FI model and obtained broadly similar results to the ones reported in the published version. This alternative approach has the disadvantage of not taking into account the economic constraints imposed by the model, and we do not pursue it here.}
 
The learning economy is specified by $p=4$ parameters, $\theta=(m_0,\gamma_{\bar{k}},b,\sigma_\delta)'$, while the FI economy is specified by $q=3$ parameters, $\phi = (m_0,\gamma_{\bar{k}},b)'.$ For this reason, the definition of the auxiliary estimator requires an additional statistic $\hat{\eta}_T\in\mathbb{R}.$ Since the noise parameter $\sigma_\delta$ controls the skewness of excess returns, the third moment seems like a natural choice. We are concerned, however, that the third moment may be too noisy to produce an efficient estimator of $\theta.$ For this reason, we consider an alternative based on the observation that by restriction (\ref{Qbar}), the mean return is nearly independent of the structural parameter: 
\begin{equation}
\mathbb{E}(r_t)\approx \ln(1+1/\overline{Q})+g_D-r_f-\overline{\sigma}^2_{D},
\label{risk premium}
\end{equation}
as is verified in the online appendix. Since the mean is fixed, the median can be used as a robust measure of skewness. The auxiliary estimator $\hat\mu_T=(\hat\phi_T,\hat\eta_T)'$ is defined by expanding the ML estimator of the full-information economy, $\hat\phi_T$, with either the third moment ($\hat\eta_T=T^{-1}\sum_{t=1}^T r_t^3$) or median ($\hat\eta_T=\text{median}\{r_t\}$) of returns.

\begin{figure}
\tv\tv
\includegraphics[width=1\textwidth]{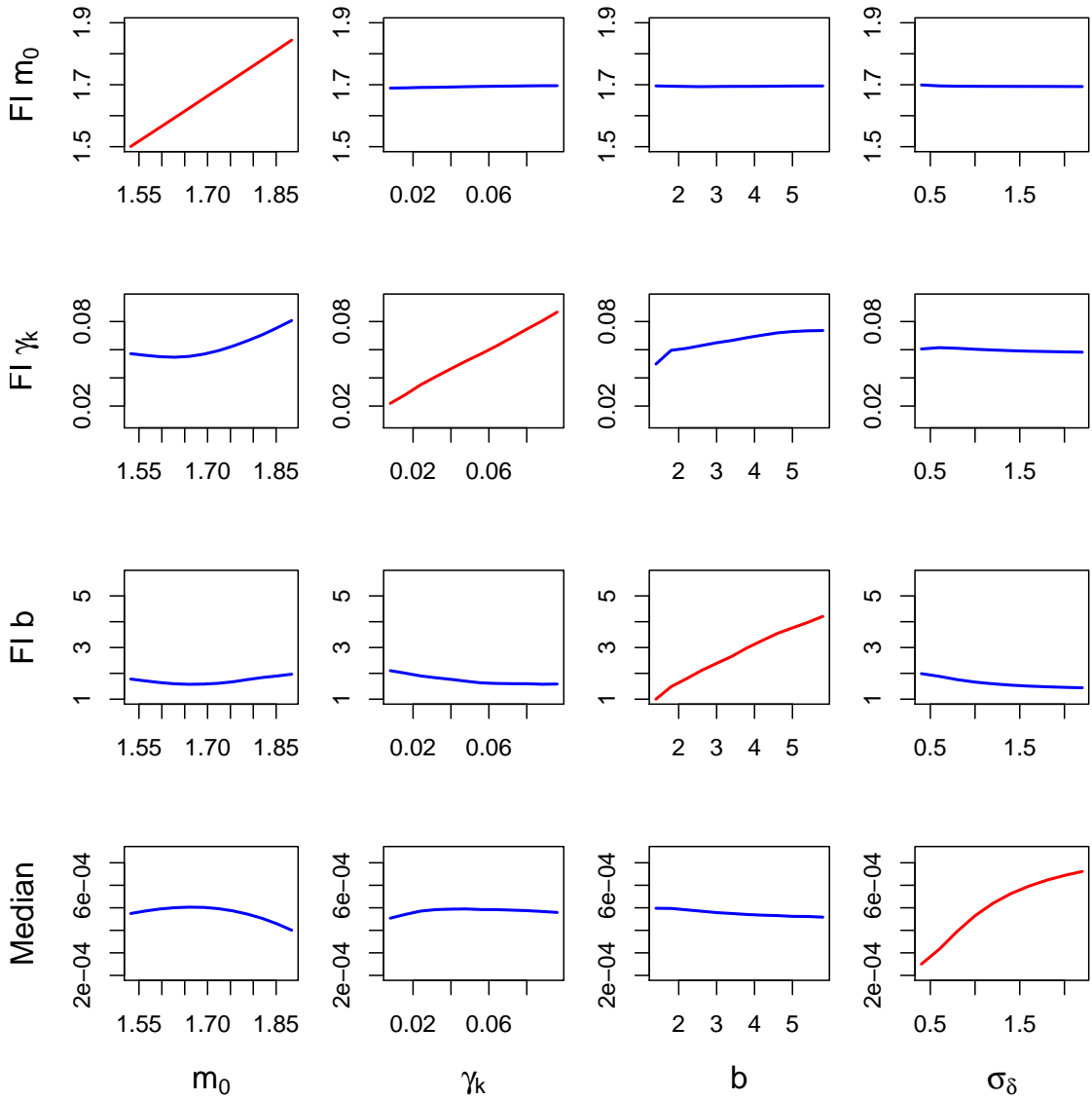}
\caption[]{Auxiliary Estimator. This figure illustrates the relation between the median-based auxiliary estimator and the structural parameter $\theta.$ In each column, one structural parameter is allowed to vary while the other three parameters are set to their reference values. The auxiliary estimate reported for every $\theta$ is obtained from a simulated sample of length $10^7$ generated under the learning model $\theta$ with $\overline{k}=3$ volatility components. }
\label{binding}
\end{figure}

In Figure~\ref{binding}, we illustrate the relation between the median-based auxiliary estimator $\hat\mu_T$ and the structural parameter $\theta$ on a long simulated sample of length $ST=10^7$. The graphs can be viewed as cuts of the binding function $\mu(\theta)$. The top three rows show that for all $i\in\{1,2,3\}$, the auxiliary parameter $\hat{\mu}_{T,i}$ increases monotonically with the corresponding parameters $\theta_i$ of the learning economy, and is much less sensitive to the other parameters $\theta_j,$ $j \neq i$ (including $\sigma_\delta$). Moreover, we note that the auxiliary estimator of $b$, based on FI ML, is a biased estimator of the parameter $b$ of the incomplete-information economy; this finding illustrates the pitfalls of employing quasi-maximum likelihood estimation in this setting. The bottom row shows that as the noise parameter $\sigma_{\delta}$ increases, the median return increases monotonically, consistent with the fact that returns become more negatively skewed. In the online appendix, we verify that the third moment is decreasing monotonically with $\sigma_\delta.$ The structural parameter $\theta$ is thus well identified by our two candidate auxiliary estimators.
 
As a benchmark, we also construct a simulated method of moments (SMM) estimator. In the online appendix, we illustrate the impact of the structural parameter $\theta$ on the expected values of $r^{n}_{t}$, $n\in\{1,\dots,4\}$, the leverage coefficient $r_{t-1}r_{t}^2$, and the volatility autocorrelation measure $r^{2}_{t-1}r^{2}_{t}$. The leverage measure and the second, third and fourth moments appear to be the most sensitive to the structural parameter $\theta,$ and are therefore selected for the definition of the SMM estimator.
  
\begin{figure}
\includegraphics[height=0.6\textheight]{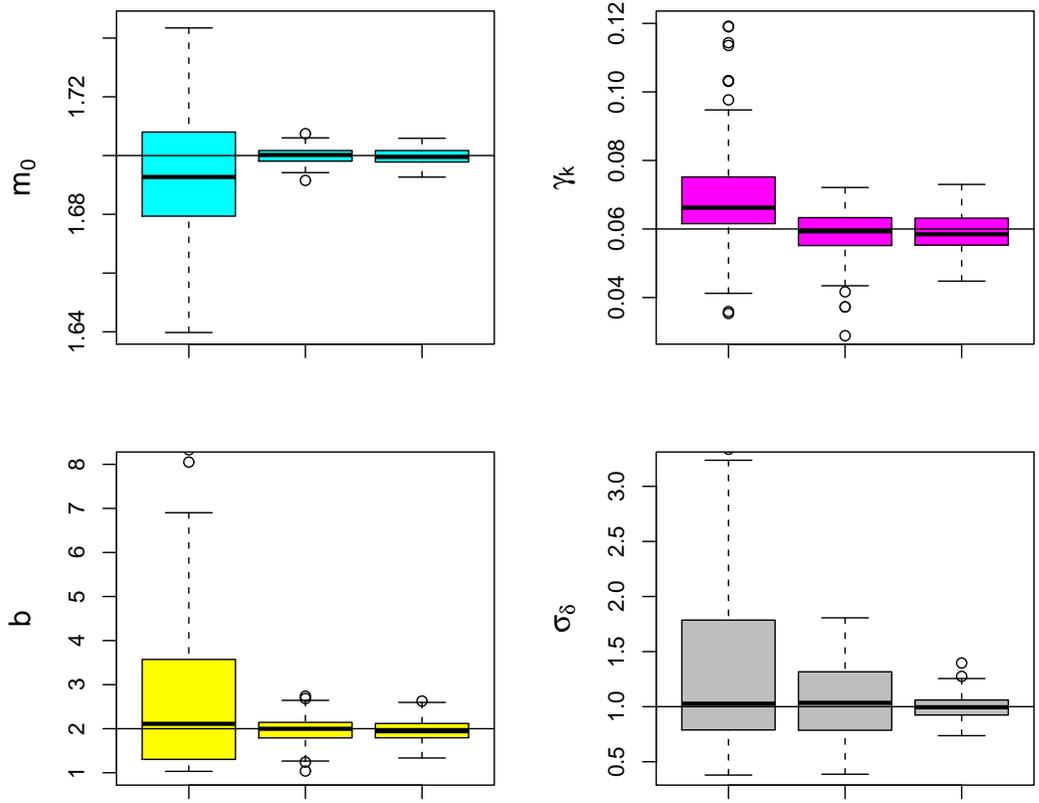} 
\caption[]{Monte Carlo Simulations of the Learning Model Estimators. This figure illustrates boxplots of the structural parameter estimates obtained using SMM (left boxplot of each panel), the indirect inference estimator based on the third moment (middle boxplots), and the median-based indirect inference estimator (right boxplots). The horizontal lines correspond to the true value of each parameter.}
\label{boxplots}
\end{figure}
 
In Figure~\ref{boxplots}, we report boxplots of SMM, third moment-based and median-based II estimates of $\theta$ obtained from 100 simulated sample paths of length $T=20,000$ from the learning model with $\bar{k}=3$ volatility components. For all three estimators, we set the simulation size to $S=500$, so that each simulated path contains $ST=10^7$ simulated data points. The indirect inference procedures provide more accurate and less biased estimates of the structural parameters of the learning economy than SMM. The median-based estimator provides substantially more accurate estimates of the parameter $\sigma_\delta$ that controls the agent's information quality. The median-based estimator thus strongly dominates the other two candidate estimators, and we now use it empirically. Overall, the Monte Carlo simulations confirm the excellent properties of the filtering and estimation techniques proposed in the paper. 

\subsection{Empirical Estimates and Value at Risk Forecasts}
\begin{figure}
\includegraphics[height=0.5\textheight]{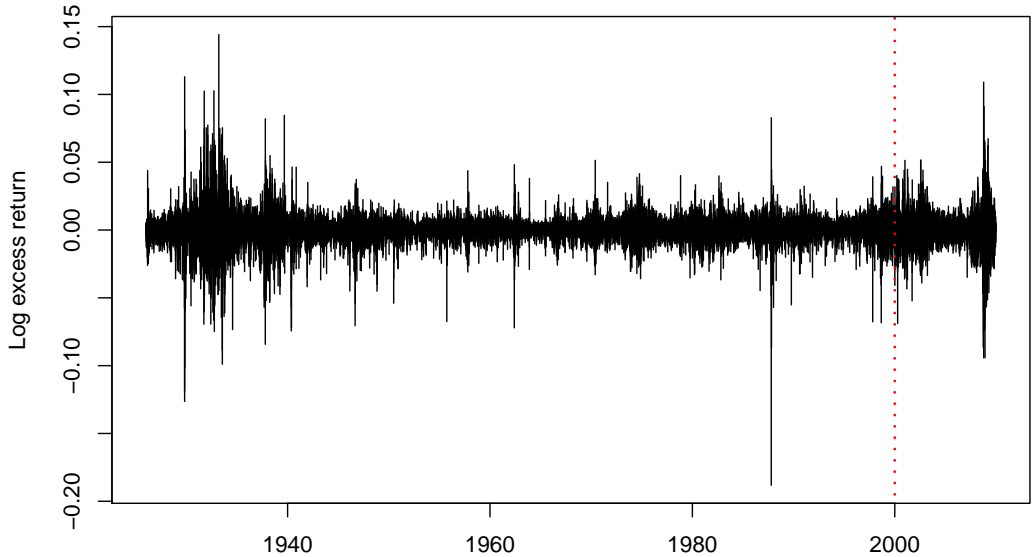}
\caption[]{U.S. Equity Return Data. This figure illustrates the daily log excess returns on the CRSP U.S. value-weighted equity index between 2 January 1926 and 31 December 2009. The dashed line separates the in-sample and out-of-sample periods.}
\label{CRSP}
\end{figure}

We apply our estimation methodology to the daily log excess returns on the U.S. CRSP value-weighted equity index from 2 January 1926 to 31 December 2009. The dataset contains 22,276 observations, which are illustrated in Figure~\ref{CRSP}. We partition the dataset into an in-sample period, which runs until 31 Dec 1999, and an out-of-sample period, which covers the remaining ten years.

\begin{table}
\begin{center}
\begin{minipage}[t]{9cm}
   \renewcommand{\footnoterule}{} 
\caption{{\sc{Empirical estimates}}\protect\footnote{We report empirical estimates of the learning model (with standard errors in parentheses) based on the daily excess returns of the CRSP index between 2 January 1926 and 31 December 1999. The log-likelihood estimates are based on an SOS filter containing $N=10^7$ particles. HAC-adjusted Vuong tests comparing $\overline{k}\leq 3$ specifications to $\overline{k}=4$ are reported in parentheses below the log-likelihood estimates.}}
\vskip5pt
\begin{tabular}{l|cccc|c}
\hline\hline
\tvi
$\overline{k}$ & \multicolumn{4}{c}{Parameter Estimates} &  \multicolumn{1}{|c}{Estimated}\\
& $m_0$ & $\gamma_{\overline{k}}$ & $b$ & $\sigma_\delta$ & Likelihood\\
& & & & & (in logs)\\
\hline
\tvi 1 & $\underset{(0.0091)}{1.732}$ & $\underset{(0.0033)}{0.063}$ & -
& $\underset{(61,616.8)}{93.807}$ & $\underset{(-10.4948)}{65,680.1}$\\
2 & $\underset{(0.0061)}{1.714}$ & $\underset{(0.0036)}{0.054}$ &
$\underset{(10.5573)}{21.104}$ & $\underset{(1.1036)}{4.001}$ & $\underset{(-8.0477)}{67,104.9}$\\
3 & $\underset{(0.0055)}{1.690}$ & $\underset{(0.0055)}{0.071}$ &
$\underset{(9.9115)}{16.471}$ & $\underset{(1.5599)}{2.401}$ & $\underset{(-8.6697)}{67,534.7}$\\
4 & $\underset{(0.0059)}{1.587}$ & $\underset{(0.0049)}{0.047}$ &
$\underset{(0.5387)}{5.089}$ & $\underset{(0.1714)}{1.411}$ & $68,167.8$\\
\hline
\end{tabular}
\label{IIin}
\end{minipage}
\end{center}
\end{table}

In Table~\ref{IIin}, we report the II estimates of $\theta.$ We let $ST=10^7$ and report standard errors in parentheses. The estimate of $\sigma_{\delta}$ is significant and declines with $\bar{k}$.\footnote{When $\bar{k}=1$, the auxiliary parameter is nearly invariant to $\sigma_\delta$ in the relevant region of the parameter space. The Jacobian of the binding function is almost singular, and by (\ref{IIvar}), the estimator of $\sigma_\delta$ has a very large asymptotic variance. The specification with $\bar{k}=1$ cannot match the median of historical returns and is therefore severely misspecified. These findings   illustrate the empirical importance of using higher values of $\bar{k}$.} This finding is consistent with the intuition that as $\overline{k}$ increases, the effect of learning becomes increasingly powerful, and a lower $\sigma_{\delta}$ better matches the negatively skewed excess return series. We also report the log-likelihood of each specification, which is estimated by an SOS filter with $N=10^7$ particles every period. The likelihood function of the II model increases steadily with $\bar{k}$. We report in parentheses the $t-$ratios of a HAC-adjusted Vuong (1989) test, that is the rescaled differences between the log-likelihoods of the lower-dimensional ($\bar{k}\in\{1,2,3\}$) and the highest-dimensional ($\bar{k}=4$) specifications. The four-component model has a significantly higher likelihood than the other specifications and is therefore selected for the out-of-sample analysis.

\begin{table}
\begin{center}
\begin{minipage}[t]{12.0cm}
   \renewcommand{\footnoterule}{} 
\caption{{\sc{Failure rates of value-at-risk forecasts}}\protect\footnote{This table reports the failure rates of the 1-day and 5-day value at risk forecasts produced by various methods in the out-of-sample period (2000-2009). The historical VaR is based on a rolling window of 60 days. The GARCH, FI and II forecasts are computed using in-sample parameter estimates. II forecasts are based on an SOS filter with $N=10^7$ elements. The significance level is $1\%$.}}
\begin{center}
\vskip5pt
\begin{tabular}{l|ccc|ccc}
\hline\hline
Models & \multicolumn{3}{c|}{One Day} & \multicolumn{3}{c}{Five Days}\\ 
 & $1\%$ & $5\%$ & $10\%$ & $1\%$ & $5\%$ & $10\%$\\
\hline
Historical VaR & $-$ & $\underset{(0.0051)}{\textbf{0.069}}$ &
$\underset{(0.0065)}{\textbf{0.119}}$ & $-$ &
$\underset{(0.0111)}{0.066}$ & $\underset{(0.0150)}{0.129}$\\ 
\hline
GARCH & $\underset{(0.0054)}{\textbf{0.081}}$ &
$\underset{(0.0072)}{\textbf{0.154}}$ & $\underset{(0.0079)}{\textbf{0.197}}$ &
$\underset{(0.0095)}{\textbf{0.048}}$ &
$\underset{(0.0147)}{\textbf{0.123}}$ &
$\underset{(0.0166)}{\textbf{0.165}}$ \\ 
\hline
$FI, \overline{k}=4$ & $\underset{(0.0025)}{0.016}$ &
$\underset{(0.0051)}{\textbf{0.070}}$ & $\underset{(0.0067)}{\textbf{0.132}}$ & 
$\underset{(0.0048)}{0.012}$ & $\underset{(0.0112)}{0.068}$ &
$\underset{(0.0156)}{\textbf{0.143}}$ \\
\hline
$II, \overline{k}=4$ & $\underset{(0.0018)}{0.008}$ &
$\underset{(0.0042)}{0.047}$ & $\underset{(0.0058)}{0.094}$ & 
$\underset{(0.0052)}{0.014}$ & $\underset{(0.0106)}{0.060}$ &
$\underset{(0.0153)}{{0.135}}$ \\ 
\hline
\end{tabular}
\end{center}
\label{VaR}
\end{minipage}
\end{center}
\end{table}

We now turn to the out-of-sample implications of the incomplete-information model. The value at risk $VaR_{t+1}^p$ constructed on day $t$ is such that the return on day $t+1$ will be lower than $-VaR_{t+1}^p$ with probability $p.$ The failure rate is specified as the fraction of observations where the actual return exceeds the value at risk. In a well specified VaR model, the failure rate is on average equal to $p.$ We use as a benchmark historical simulations (e.g. Christoffersen 2009) and Student GARCH(1,1), which are widely used in practice. The historical VaR estimates are based on a window of 60 days, which corresponds to a calendar period of about three months. 
In Table \ref{VaR}, we report the failure rates of the $VaR_{t+1}^p$ forecasts for $p=1\%,\,5\%,\,10\%$, at horizons of 1 and 5 days produced by: historical simulations, GARCH, the full-information model and the learning model with $\bar{k}=4$. Standard deviations are reported in parentheses. A failure rate is in bold characters if it differs from its theoretical value at the 1\% significance level.

Historical simulations provide inaccurate VaR forecasts at the 1-day horizon. The failure rates are significantly higher than their theoretical values, which suggests that historical simulations provide overly optimistic estimates of value at risk. GARCH VaR estimates are significantly higher in all cases, while the FI model's VaR predictions are rejected in three out of six cases. 
On the other hand, the VaR predictions from the learning model are all consistent with the data. Our empirical findings suggest that the learning model captures well the dynamics of daily stock returns, and outperforms out of sample some of the best reduced-form specifications. We note that this is an excellent result for a consumption-based asset pricing model.

\section{Conclusion}
In this paper, we have developed powerful filtering and estimation methods for a wide class of learning environments. The new SOS algorithm applies to general state space models in which state-observation pairs can be conveniently simulated. Our method makes no assumption on the availability of the observation density and therefore expands the scope of sequential Monte Carlo methods. The rate of convergence does not depend on the size of the state space, which shows that our filter defeats a form of the curse of dimensionality. Among many possible applications, SOS is useful to estimate the likelihood function, conduct likelihood-based specification tests, and generate forecasts. 

The new filter naturally applies to nonlinear economies with agent learning of the type often considered in financial economics. In this context, SOS permits to track in real time both fundamentals and agent beliefs about fundamentals. Estimation can proceed by simulated ML, but this approach can be computationally costly, as in the example of section 4. For this reason, we have defined an indirect inference estimator by expanding the full-information MLE with a set of statistics that agent learning is designed to capture. 

These methods have been applied to a consumption-based asset pricing model with investor learning about multifrequency volatility. We have verified by Monte Carlo simulations the accuracy of our SOS filter and indirect inference estimators, and have implemented these techniques on a long series of daily excess stock returns. We have estimated the parameters driving fundamentals and the quality of the signals received by investors, tracked fundamentals and investor beliefs over time, and verified that the inferred specification provides good value-at-risk forecasts out of sample. 

The paper opens multiple directions for future research. SOS can be used to price complex instruments, such as derivatives contracts, which crucially depend on the distribution of the hidden state. We can expand the role of learning in the analysis, for instance by letting the agent learn the parameter of the economy over time, or by conducting the joint online estimation of the structural parameter $\theta$ and the state of the economy $s_t$, as in Polson, Stroud, and Mueller (2008) and Storvik (2002). Further extensions could include inference for equilibrium models with asymmetric information (e.g. Biais, Bossaerts, and Spatt, 2010), and the development of value-at-risk models that incorporate the cross-sectional dispersion of investor beliefs.

\newpage
\appendix
 
\section{Convergence of the SOS Filter (Section~\ref{SOS})}\label{app}
\subsection{A Preliminary Result}
In this appendix, we show the convergence of the SOS particle filter defined in section 2 as the number of particles $N$ goes to infinity. Since the path $R_T$ is fixed, our focus is on simulation noise, and expectations in this section are over all the realizations of the random particle method. We begin by establishing the following result for a given $N\geq1$ and $t\geq1$.
\vskip10pt

\noindent\textbf{Lemma A1.} \textit{Assume that there exists $U_{t-1}(N)$ such that for every bounded measurable function $\Phi:\mathcal{S}\rightarrow\mathbb{R},$ 
\begin{equation}
\mathbb{E}\left\{\left[\frac{1}{N}\sum_{n=1}^N \Phi(s_{t-1}^{(n)}) - \mathbb{E}[\Phi(s_{t-1})|R_{t-1}]\right]^2\right\} \leq U_{t-1}(N)\|\Phi\|^2.
\label{condition_A1}
\end{equation}
Let $U^*_t(N) = 2\kappa'^2_t A(K)^2 h_t^4+ B(K)\kappa_t/(Nh_t^{n_R})
+2U_{t-1}(N)\kappa_t^2.$ Then, the inequality
\begin{equation*}
\mathbb{E}\left\{\left[\frac{1}{N}\sum_{n=1}^N \Phi(\tilde{s}_t^{(n)}) K_{h_t}(r_t - \tilde{r}_t^{(n)})-f_R(r_t|R_{t-1}) \mathbb{E}\left[\Phi(\left.s_t)\right|R_t\right]\right]^2\right\}\leq U^*_t(N) \|\Phi\|^2 
\end{equation*}
holds for every bounded measurable function $\Phi$.}

\vskip10pt
\noindent\textbf{Proof of Lemma A1.} 
We consider the function 
\begin{equation*}
a_{t-1}(s_{t-1}) = \int{\Phi(\tilde{s}_t)K_{h_t}(r_t-\tilde{r}_t)g(d\tilde{s}_t,d\tilde{r}_t|s_{t-1},R_{t-1})}.
\end{equation*}
We note that 
\begin{equation*}
\begin{aligned}
|a_{t-1}(s_{t-1})|
& \leq
\|\Phi\|\int{K_{h_t}(r_t-\tilde{r}_t)g(d\tilde{s}_t,d\tilde{r}_t|s_{t-1},R_{t-1})}\\
&= \|\Phi\|\int{K_{h_t}(r_t-\tilde{r}_t)f_R(\tilde{r}_t|s_{t-1},R_{t-1})d\tilde{r}_t.}
\end{aligned}
\end{equation*}
The function $a_{t-1}$ is therefore bounded above by $\kappa_t\hskip3pt\|\Phi\|$.
 
The difference
$Z = N^{-1}\sum_{n=1}^N \Phi(\tilde{s}_t^{(n)}) K_{h_t}(r_t - \tilde{r}_t^{(n)}) - f_R(r_t|R_{t-1}) \mathbb{E}\left[\Phi(\left.s_t)\right|R_t\right]$
is the sum of the following three terms:
\begin{equation*}
\begin{aligned} 
&Z_1 = \frac{1}{N}\sum_{n=1}^N \left[\Phi(\tilde{s}_t^{(n)})K_{h_t}(r_t-\tilde{r}_t^{(n)}) - a_{t-1}(s_{t-1}^{(n)})\right], \\
&Z_2 = \frac{1}{N}\sum_{n=1}^N a_{t-1}(s_{t-1}^{(n)}) - \int{a_{t-1}(s_{t-1})\lambda(ds_{t-1}|R_{t-1})},\\
&Z_3 = \int{a_{t-1}(s_{t-1})\lambda(ds_{t-1}|R_{t-1})} -f_R(r_t|R_{t-1})\mathbb{E}\left[\Phi(\left.s_t)\right|R_t\right].
\end{aligned}
\end{equation*}
Let $\mathcal{S}_{t-1}^{(N)}=(s_{t-1}^{(1)},\dots,s_{t-1}^{(N)})$ denote the vector of period$-(t-1)$ particles. Conditional on $\mathcal{S}_{t-1}^{(N)},$ $Z_1$ has a zero mean, while $Z_2$ and $Z_3$ are deterministic. Hence:
\begin{equation*}
\mathbb{E}(Z^2) = \mathbb{E}(Z_1^2) + \mathbb{E}[(Z_2+Z_3)^2] \leq \mathbb{E}(Z_1^2) + 2\mathbb{E}(Z_2^2)+ 2\mathbb{E}(Z_3^2).
\end{equation*}
Conditional on $\mathcal{S}_{t-1}^{(N)},$ the state-observation pairs $\{(\tilde{s}_t^{(n)},\tilde{r}_t^{(n)})\}_{n=1}^N$ are independent, and each $(\tilde{s}_t^{(n)},\tilde{r}_t^{(n)})$ is drawn from  $g(\cdot|s_{t-1}^{(n)},R_{t-1});$ the addends of $\Phi(\tilde{s}_t^{(n)})K_{h_t}(r_t-\tilde{r}_t^{(n)}) - a_{t-1}(s_{t-1}^{(n)})$ are thus independent and have mean zero. We infer that the conditional expectation of $Z_1^2$ is bounded above by:
\begin{equation*}
\frac{1}{N^2}\sum_{n=1}^N\int{\Phi(\tilde{s}_t)^2K_{h_t}(r_t-\tilde{r}_t)^2  g(d\tilde{s}_t,d\tilde{r}_t|s_{t-1}^{(n)},R_{t-1})}
\leq \frac{\kappa_t\|\Phi\|^2}{N} \int{K_{h_t}(r_t-\tilde{r}_t)^2 d\tilde{r}_t}.
\end{equation*}
We apply the change of variable $u=(r_t-\tilde{r}_t)/h_t$:
\begin{equation*}
\int{K_{h_t}(r_t-\tilde{r}_t)^2 d\tilde{r}_t}
= \frac{B(K)}{h_t^{n_R}},
\end{equation*}
and infer that $\mathbb{E}(Z_1^2) \leq \|\Phi\|^2 B(K) \kappa_t/(Nh_t^{n_R}).$
 
Since the function $a_{t-1}(s_{t-1})$ is bounded above by $\kappa_t\hskip3pt \|\Phi\|,$ we infer from (\ref{condition_A1}) that:
$\mathbb{E}(Z_2^2)\leq U_{t-1}(N)\kappa_t^2 \hskip3pt \|\Phi\|^2.$
 
Finally, we observe that $ f_R(r_t|R_{t-1}) \mathbb{E}\left[\Phi(\left.s_t)\right|R_t\right]=\int{\Phi(s_t)f_R(r_t|s_t,R_{t-1})\lambda(ds_t|R_{t-1})},$ and therefore
\begin{equation*}
\begin{aligned}
Z_3 
&= \int{\Phi(s_t)\left\{\int{K_{h_t}(r_t-\tilde{r}_t)[f_R(\tilde{r}_t|s_t,R_{t-1})-f_R(r_t|s_t,R_{t-1})]d\tilde{r}_t}\right\}\lambda(ds_t|R_{t-1})}\\
&=\int{\Phi(s_t)\left\{\int{K(u)[f_R(r_t-h_t u|s_t,R_{t-1})-f_R(r_t|s_t,R_{t-1})]du}\right\}\lambda(ds_t|R_{t-1})}.
\end{aligned}
\end{equation*}
Note that $\left|\int{K(u)[f_R(r_t-h_tu|s_t,R_{t-1})-f_R(r_t|s_t,R_{t-1})]du}\right| \leq \kappa'_t  A(K) h_t^2.$
Hence $|Z_3|\leq \kappa'_t A(K) h_t^2 \|\Phi\|$ and therefore
$\mathbb{E}(Z^2_3)\leq \kappa_t'^2 A(K)^2 h_t^4\|\Phi\|^2.$ We conclude that the lemma holds.\hfill{\textit{Q.E.D.}}

\subsection{Proof of Theorem~\ref{theo}}
The proof of (\ref{MSEt}) proceeds by induction. When $t=0,$ the particles are drawn from the prior $\lambda_0$, and the conditional expectation is computed under the same prior. Hence the property (\ref{MSEt}) holds with $U_0(N)= 1/N.$
 
We now assume that the property (\ref{MSEt}) holds at date $t-1$. The estimation error
$X = N^{-1}\sum_{n=1}^N \Phi(s_t^{(n)}) - \mathbb{E}[\Phi(s_t)|R_t]$ is the sum of:
\begin{equation*}
\begin{aligned}
&X_1 = \frac{1}{N}\sum_{n=1}^N \Phi(s_t^{(n)}) -
\sum_{n=1}^N p^{(n)}_t \Phi(\tilde{s}_t^{(n)}) .\\
&X_2 = \left[\sum_{n=1}^N p^{(n)}_t \Phi(\tilde{s}_t^{(n)})\right] \left[\frac{f_R(r_t|R_{t-1}) - N^{-1}\sum_{n'=1}^N K_{h_t}(r_t-\tilde{r}_t^{(n')})}{f_R(r_t|R_{t-1})}\right], \\
&X_3 = \frac{1}{Nf_R(r_t|R_{t-1})}\sum_{n=1}^N \Phi(\tilde{s}_t^{(n)}) K_{h_t}(r_t-\tilde{r}_t^{(n)}) - \mathbb{E}[\Phi(s_t)|R_t].
\end{aligned}
\end{equation*}
The first term, $X_1$, corresponds to step 3 resampling, the second term to the normalization of the resampling weights, and the third term to the error in the estimation of $\Phi$ using the nonnormalized weights.
 
Conditional on $\{(\tilde{s}^{(n)}_t,\tilde{r}^{(n)}_t)\}_{n=1}^N,$ the particles $s_t^{(n)}$ are independent and identically distributed, and $X_1$ has mean zero. We infer that 
$\mathbb{E}[X_1^2|\{\tilde{s}_t^{(n)},\tilde{r}_t^{(n)}\}_{n=1}^N] \leq \|\Phi\|^2/N,$
and therefore $\mathbb{E}(X_1^2)\leq \|\Phi\|^2/N.$ Note that when we use stratified, residual or combined stratified-residual resampling in step 3, the inequality $\mathbb{E}(X_1^2)\leq \|\Phi\|^2/N$ remains valid, and smaller upper bounds can also be derived.\footnote{See Capp\'e, O., Moulines, E., and T. Ryd\'en (2005, ch. 7) for a detailed discussion of sampling variance.}
 
Conditional on $\{(\tilde{s}^{(n)}_t,\tilde{r}^{(n)}_t)\}_{n=1}^N,$ $X_2$ and $X_3$ are deterministic variables. The mean squared error satisfies: 
\begin{equation*}
\mathbb{E}(X^2) = \mathbb{E}(X_1^2) + \mathbb{E}[(X_2+X_3)^2] \leq \mathbb{E}(X_1^2) + 2\mathbb{E}(X_2^2)+2\mathbb{E}(X_3^2).
\end{equation*} 
 
We note that $|X_2| \leq \|\Phi\|[f_R(r_t|R_{t-1})]^{-1}\left|f_R(r_t|R_{t-1}) - \sum_{n'=1}^N K_{h_t}(r_t-\tilde{r}_t^{(n')})/N\right|.$ Using the induction hypothesis at date $t-1$, we apply Lemma A1 with $\Phi \equiv 1$ and obtain that $\mathbb{E}(X_2^2)$ is bounded above by:
\begin{equation}
\frac{U_t^*(N)\|\Phi\|^2}{[f_R(r_t|R_{t-1})]^2}.
\label{upper_bound}
\end{equation}
Lemma A1 implies that $\mathbb{E}(X_3^2)$ is also bounded above by (\ref{upper_bound}). We conclude that $\mathbb{E}(X^2) \leq U_t(N)\|\Phi\|^2,$
where $U_t(N)=4 U_t^*(N)[f_R(r_t|R_{t-1})]^{-2} + N^{-1}$, or equivalently
\begin{equation}
U_t(N)=\frac{4}{[f_R(r_t|R_{t-1})]^2} \left[2\kappa'^2_t A(K)^2 h_t^4+ \frac{B(K)\kappa_t}{Nh_t^{n_R}}+2U_{t-1}(N)\kappa_t^2\right] + \frac{1}{N}.
\label{recursive U}
\end{equation}
This establishes part (\ref{MSEt}) of the theorem. From (\ref{MSEt}) and Lemma~A1 with $\Phi\equiv 1$, (\ref{likelihood_bound}) follows.  
 
Assume now that the bandwidth is a function of $N$, and that assumption \ref{bandwidth} holds. A simple recursion implies that $\lim_{N\rightarrow\infty}U_t(N)=0$ for all $t.$ The mean squared error converges to zero for any bounded measurable function $\Phi.$
 
We now characterize the rate of convergence. Given $U_{t-1}(N)$, we know that the coefficient $U_t(N)$ defined by (\ref{recursive U}) is minimal if 
\begin{equation}
h_t = N^{-1/(n_R+4)}\left[\frac{\kappa_t n_RB(K)}{8 \kappa'^2_t A(K)^2}\right]^{1/(n_R+4)}.
\end{equation}
More generally, if the bandwidth sequence is of the form $h_t(N) = h_t(1)/N^{-1/(n_R+4)}$, then $U_t(N)$ is of the form:
\begin{equation}
U_t(N)=u_{1,t}N^{-4/(n_R+4)} + u_{2,t} U_{t-1}(N) + N^{-1}.
\end{equation}
where $u_{1,t}$ and $u_{2,t}$ are finite nonnegative coefficients.\footnote{We verify that $u_{1,t}=4[f(r_t|R_{t-1})]^{-2}\left[2\kappa'^2_t h_t(1)^4 A(K)^2 + B(K)\kappa_t h_t(1)^{-n_R}\right]$ and $u_{2,t}=8\kappa_t^2[f(r_t|R_{t-1})]^{-2}.$} By a simple recursion, $U_t(N)$ is of order $N^{-4/(n_R+4)}$ for all $t.$\hfill{\textit{Q.E.D.}}

\vskip20pt
\section{Learning Economies (Section~\ref{learning})}
 
\subsection{Proof of Proposition~\ref{AgentBelief}} 
We infer from Bayes' rule that
\begin{equation*}
\Pi_t^j \propto \underset{=f_X(x_t|M_t=m^j;\theta){\text{ by As.~\ref{signal}(b)}}}{\underbrace{f_X(x_t|M_t=m^j,X_{t-1};\theta)}} \mathbb{P}(M_t=m^j|X_{t-1};\theta),
\end{equation*}
where 
\begin{equation*}
\mathbb{P}(M_t=m^j|X_{t-1};\theta)=\sum_{i=1}^d \underset{=a_{ij}{\text{ by As.~\ref{signal}(a)}}}{\underbrace{\mathbb{P}(M_t=m^j|M_{t-1}=m^i,X_{t-1};\theta)}} \mathbb{P}(M_{t-1}=m^i|X_{t-1};\theta),
\end{equation*}
and Proposition~\ref{AgentBelief} holds.\hfill{\textit{Q.E.D.}}

\subsection{Proof of Proposition~\ref{ergodic}}
Bayes' rule (\ref{Bayes}) implies that for every $i\in\{1,\dots,d\},$ 
\begin{equation}
\Pi_t|M_t=m^i,s_{t-1},\dots,s_1\,\sim\,\Pi_t|M_t=m^i,\Pi_{t-1}\,.
\label{erg1}
\end{equation}
Also, by Assumption~\ref{signal}(a)
\begin{equation}
\mathbb{P}(M_t =m^i|s_{t-1},\dots,s_1;\theta)=\mathbb{P}(M_t =m^i|M_{t-1};\theta)\,.
\label{erg2}
\end{equation}
From (\ref{erg1}) and (\ref{erg2}), we conclude that $s_t$ is first-order Markov.
 
We know from Kaijser (1975) that under the conditions stated in the proposition, the belief process $\Pi_t$ has a unique invariant distribution. Proposition 2.1 in van Handel (2009) implies that $(M_t,\Pi_t)$ also has a unique invariant measure $\Lambda_\infty.$\footnote{Chigansky (2006) derives a similar result in continuous time.} We infer from the Birkhoff-Khinchin theorem that for any integrable function $\Phi:\mathcal{S}\rightarrow\mathbb{R},$ the sample average
$T^{-1}\sum_{t=1}^T \Phi(s_t)$ converges almost surely to the expectation of $\Phi$ under the invariant measure $\Lambda_\infty.$\hfill{\textit{Q.E.D.}}

\subsection{Proof of Proposition~\ref{closedformL}}
The econometrician recursively applies Bayes'rule:
\begin{equation*}
\mathbb{P}(M_t=m^j|R_t;\phi) = \frac{ f_{R,FI}(r_t|M_t = m^j,R_{t-1};\phi) \mathbb{P}(M_t = m^j|R_{t-1};\phi)}{f_{R,FI}(r_t|R_{t-1};\phi)}
\end{equation*}
Since $f_{R,FI}(r_t|M_t = m^j,R_{t-1};\phi) = \sum_{i=1}^d f_{i,j}(r_t;\phi) \mathbb{P}(M_{t-1}=m^i|M_t = m^j,R_{t-1};\phi),$
we infer that 
$f_{R,FI}(r_t|M_t = m^j,R_{t-1};\phi) \mathbb{P}(M_t=m^j|R_{t-1};\phi) = \sum_{i=1}^d f_{i,j}(r_t;\phi) \mathbb{P}(M_{t-1}=m^i,M_t = m^j|R_{t-1};\phi),$ and therefore
\begin{equation*}
\mathbb{P}(M_t=m^j|R_t;\phi) = 
\frac{ \sum_{i=1}^d a_{i,j} f_{i,j}(r_t;\phi)\mathbb{P}(M_{t-1}=m^i|R_{t-1};\phi)}{f_{R,FI}(r_t|R_{t-1};\phi)}.
\end{equation*}
The econometrician's conditional probabilities are therefore computed recursively.
 
Since the conditional probabilities $\mathbb{P}(M_t=m^j|R_t;\phi)$ add up to unity, the conditional density of $r_t$ satisfies
\begin{equation*}
f_{R,FI}(r_t|R_{t-1};\phi)=\sum_{i=1}^d \sum_{j=1}^d a_{i,j} f_{i,j}(r_t;\phi)\mathbb{P}(M_{t-1}=m^i|R_{t-1};\phi).
\end{equation*}
The log-likelihood function $\mathcal{L}_{FI}(\phi|R_T)=\sum_{t=1}^T \ln f_{R,FI}(r_t|R_{t-1};\phi)$ thus has an analytical expression.\hfill{\textit{Q.E.D.}}

\subsection{Indirect Inference Estimator}
We provide a set of sufficient conditions for the asymptotic results at the end of section 3.2, and then discuss numerical implementation.

\subsubsection{Sufficient Conditions for Convergence}
We assume that $\hat{\eta}_{T}$ maximizes a criterion $\mathcal{H}(\eta,R_T)$ that does not depend on the full-information MLE $\hat{\phi}_T.$ The auxiliary estimator $\hat{\mu}_T=(\hat{\phi}'_T,\hat{\eta}'_{T})'$ can therefore be written as:
\begin{equation}
\label{criterion}
\hat\mu_T={\text{arg}} \max_\mu \mathcal{Q}_T(\mu,R_T),
\end{equation} 
where $\mathcal{Q}_T(\mu,R_T)=T^{-1}\mathcal{L}_{FI}(\phi,R_T) + \mathcal{H}(\eta,R_T)$ for all $\mu=(\phi',\eta')'.$ 

\vskip5pt

\begin{assumption}[Binding Function] Under the structural model $\theta^*,$ the auxiliary criterion function $\mathcal{Q}_T(\mu,R_T)$ converges in probability to $\mathcal{Q}_\infty(\mu,\theta^*)$ for all $\mu.$
Moreover, the function $\mu:\mathbb{R}^p\rightarrow\mathbb{R}^p$ defined by
\begin{equation*}
\mu(\theta)=\arg\,\max_{\mu}\,\mathcal{Q}_{\infty}(\mu,\theta),
\end{equation*}
called the binding function, is injective.
\label{binding function}
\end{assumption}

\vskip5pt 

\begin{assumption}[Score] The renormalized score satisfies:
\begin{equation*}
\sqrt{T}\,\frac{\partial \mathcal{Q}_T}{\partial\mu} [\mu(\theta^*),R_T] \overset{d}\rightarrow \mathcal{N} (0,I_0),
\end{equation*}
where $I_0$ is positive definite symmetric matrix.
\end{assumption}

\vskip5pt
 
\begin{assumption}[Hessian of Criterion Function] The Hessian matrix
\begin{equation*} 
-\frac{\partial^2 \mathcal{Q}_T}{\partial \mu \partial \mu'} \left[\mu(\theta^*),R_T\right]
\end{equation*}
is invertible and converges in probability to a nonsingular matrix $J_0.$
\label{Hessian}
\end{assumption}

\vskip5pt

\noindent Under assumptions \ref{binding function}-\ref{Hessian}, the auxiliary estimator satisfies
$\sqrt{T}\left[\hat{\mu}_T-\mu(\theta^*)\right]\overset{d}{\longrightarrow} \mathcal{N}(0,W^*),$
where $W^* = J_0^{-1} I_0 J_0^{-1},$ and the asymptotic results at the end of section~\ref{IIsec} hold (Gouri\'eroux, Monfort, and Renault, 1993; Gouri\'eroux and Monfort, 1996).

\subsubsection{Numerical Implementation}
Since in the just-identified case $\hat\mu_{ST}(\hat{\theta}_T)=\hat{\mu}_T$,
the simulated auxiliary estimator $\hat\mu_{ST}(\theta)$ satisfies
\begin{equation*}
\frac{\partial Q_{ST}}{\partial\mu} [\underset{\hat\mu_T}{\underbrace{\hat\mu_{ST}(\theta)}},R_{ST}(\theta)]=0\,.
\end{equation*}
Hence, the indirect inference estimator $\hat{\theta}_T$ minimizes the EMM-type objective function:
\begin{equation}
 \left\{ \frac{\partial \mathcal{Q}_{ST}}{\partial \mu} [\hat\mu_T,R_{ST}(\theta)]\right\}' W_T 
 \left\{\frac{\partial \mathcal{Q}_{ST}}{\partial \mu} [\hat\mu_T,R_{ST}(\theta)]\right\},
\label{EMMdef}
\end{equation}
where $W_T$ is any positive-definite weighting matrix. This property can be used to compute $\hat{\theta}_T.$ For each iteration of $\theta$, the evaluation of the EMM objective function (\ref{EMMdef}) requires only the evaluation of the score. By contrast, the evaluation of the objective function (\ref{II}) requires the optimization of the FI likelihood in order to obtain $\hat{\mu}_{ST}(\theta)$. The computational advantage of EMM is substantial in applications where the calculation of the full-information MLE is expensive. 
 
In the just-identified case and under assumptions~\ref{binding function}-\ref{Mest}, the asymptotic variance-covariance matrix of the indirect inference estimator simplifies to
\begin{equation*}
\Sigma = \left(1+\frac{1}{S}\right)
     \left\{\frac{\partial^2 \mathcal{Q}_{\infty}}{\partial\theta\partial\mu'}[\mu(\theta^*),\theta^*] \,I_0^{-1}\, 
     \frac{\partial^2 \mathcal{Q}_{\infty}}{\partial\mu\partial\theta'}[\mu(\theta^*),\theta^*] \right\}^{-1}\,,
\end{equation*}
as shown in Gouri\'eroux and Monfort (1996). Note that the choice of the weighting matrix $W_T$ does not affect the asymptotic variance of the indirect inference estimator in the exactly identified case. 
 
In practice, we can estimate $I_0$ and $\frac{\partial^2\mathcal{Q}_{\infty}}{\partial\theta\partial\mu'}[\mu(\theta^*),\theta^*]$ in the following way. 

\begin{assumption}[Decomposable Score] 
The score function can be written as:
\begin{equation*}
\frac{\partial \mathcal{Q}_T}{\partial \mu}(\mu,R_T) \equiv \frac{1}{T} \sum_{t=1}^{T}\psi(r_t|R_{t-1};\mu)
\end{equation*}
for all $R_T$ and $\mu.$
\label{Mest}
\end{assumption}

\noindent Note that Assumption~\ref{Mest} is satisfied by the median-based and the third moment-based indirect inference estimators considered in section 4.
 
By Assumption~\ref{Mest}, the auxiliary parameter satisfies the first-order condition:
\begin{equation}
\frac{\partial \mathcal{Q}_T}{\partial \mu}(\hat\mu_T,R_T) = \frac{1}{T} \sum_{t=1}^{T}\psi(r_t|R_{t-1};\hat\mu_T)=0.
\label{FOC}
\end{equation}
We estimate $I_0$ by the Newey and West (1987) variance-covariance matrix:
\begin{equation}
\hat{I}_0 = \hat{\Gamma}_0 + \sum_{v=1}^\tau\left(1 - \frac{v}{\tau+1}\right)\left( \hat{\Gamma}_v + \hat{\Gamma}'_v  \right) ,
\label{newey}
\end{equation}
where $\hat{\Gamma}_v = T^{-1} \sum_{t=v+1}^{T} \psi(r_t|R_{t-1};\hat\mu_T)\psi(r_t|R_{t-1};\hat\mu_T)'.$ All the results reported in the paper are based on $\tau=10$ lags. We approximate $\frac{\partial^2\mathcal{Q}_{\infty}}{\partial\theta\partial\mu'}[\mu(\theta^*),\theta^*]$ by
\begin{equation*}
\frac{\partial^2\mathcal{Q}_{ST}}{\partial\theta\partial\mu'}[\hat\mu_T,R_{ST}(\hat{\theta}_{T})],
\end{equation*}
and obtain a finite-sample estimate of the asymptotic variance-covariance matrix $\Sigma$.

\end{document}